\documentclass[12pt]{elsarticle}

\usepackage{lineno}
\usepackage{amssymb, amsmath, bm, mathtools}
\usepackage{gensymb}
\usepackage{siunitx}
\usepackage{graphicx}
\usepackage{pgf}
\usepackage{pgfpages}
\usepackage{tikz}
\usepackage{ifthen}
\usepackage{hyperref}
\hypersetup{
    colorlinks=true,
    citecolor=cyan,
    linkcolor=blue,
    filecolor=blue,      
    urlcolor=blue,
}

\usepackage{cleveref}
\usepackage{multirow}
\usepackage[version=4]{mhchem}
\usepackage{tcolorbox}
\newtcolorbox{mybox}[1]{fonttitle=\bfseries,title=#1}
\usepackage{algorithm}
\usepackage[noend]{algpseudocode}

\usepackage{setspace}

\usepackage[a4paper, left=2.5cm, right=2.5cm, top=3cm, bottom=3cm]{geometry}

\usepackage{xcolor} %
\usepackage{enumerate} %
\usepackage{eurosym} %
\usepackage[Export]{adjustbox} %
\adjustboxset{max size={0.9\linewidth}{0.9\paperheight}}
\usepackage{mathrsfs}
\usepackage{mwe}
\usepackage[textsize=\footnotesize]{todonotes}

\definecolor{blendedred}{rgb}{0.80, 0.1, 0.3}

\usepackage{tcolorbox}

\usepackage{xcolor}
\usepackage{tikz}
\usetikzlibrary{backgrounds}
\usetikzlibrary{arrows,shapes}
\usetikzlibrary{tikzmark}
\usetikzlibrary{calc}

\usepackage{amsmath}
\usepackage{amsthm}
\usepackage{amssymb}
\usepackage{mathtools, nccmath}
\usepackage{wrapfig}
\usepackage{comment}

\usepackage{graphicx}

\usepackage{array}
\usepackage{ragged2e}
\newcolumntype{P}[1]{>{\RaggedRight\hspace{0pt}}p{#1}}
\newcolumntype{X}[1]{>{\RaggedRight\hspace*{0pt}}p{#1}}

\usepackage{tcolorbox}

\usetikzlibrary{arrows,shapes,positioning,shadows,trees,mindmap}
\usepackage[edges]{forest}
\usetikzlibrary{arrows.meta}
\colorlet{linecol}{black!75}
\usepackage{xkcdcolors} %

\usetikzlibrary{backgrounds}
\usetikzlibrary{arrows,shapes}
\usetikzlibrary{tikzmark}
\usetikzlibrary{calc}

\newcommand{\tensor}[1]{\boldsymbol{#1}}
\newcommand{\ie}{{\textit{i.e.}}}
\newcommand{\rve}{\Omega}

\renewcommand{\vector}[1]{\boldsymbol{#1}}%

\newcommand{\fft}[1]{\widehat{#1}(\boldsymbol{ \xi})}
\newcommand{\ifft}[1]{\mathcal{F}^{-1}\{#1\}}
\newcommand{\wavevector}{\vector{\xi}}

\let\oldequation\equation
\let\oldendequation\endequation

\renewenvironment{equation}
  {\linenomathNonumbers\oldequation}
  {\oldendequation\endlinenomath}

\newcommand{\greenstrain}{\tensor{E}}
\newcommand{\deformation}{\tensor{F}}

\newcommand{\smallstrain}{\tensor{\varepsilon}}

\newcommand{\stress}{\tensor{\sigma}}
\newcommand{\strain}{\tensor{e}}
\newcommand{\cauchy}{\tensor{\tau}}
\newcommand{\pkI}{\tensor{P}}

\newcommand{\macro}[1]{\overline{#1}}

\newcommand{\dV}[1]{\mathrm{d}#1}
\newcommand{\trace}[1]{\mathrm{tr}(#1)}

\newcommand{\shearmodulus}{\mu}

\newcommand{\lame}{\lambda}
\usepackage[frozencache,cachedir=.,newfloat]{minted}
\usepackage{caption}

\SetupFloatingEnvironment{listing}{name=Listing}

\graphicspath{{./}}

\journal{}

\begin{document}

\begin{frontmatter}
  \title{Simplifying FFT-based methods for solid mechanics with automatic differentiation}

\address[ifbaddress]{Institute for Building Materials, ETH Zurich, Switzerland}
\author[ifbaddress]{Mohit Pundir}
\ead{mpundir@ethz.ch} 
\author[ifbaddress]{David S. Kammer \corref{cor1}} \ead{dkammer@ethz.ch}
\cortext[cor1]{Corresponding author} 
\begin{abstract}
Fast-Fourier Transform (FFT) methods have been widely used in solid mechanics to address complex homogenization problems. However, current FFT-based methods face challenges that limit their applicability to intricate material models or complex mechanical problems. These challenges include the manual implementation of constitutive laws and the use of computationally expensive and complex algorithms to couple microscale mechanisms to macroscale material behavior. Here, we incorporate automatic differentiation (AD) within the FFT framework to mitigate these challenges. We demonstrate that AD-enhanced FFT-based methods can derive stress and tangent stiffness directly from energy density functionals, facilitating the extension of FFT-based methods to more intricate material models. Additionally,  automatic differentiation simplifies the calculation of homogenized tangent stiffness for microstructures with complex architectures and constitutive properties. This enhancement renders current FFT-based methods more modular, enabling them to tackle homogenization in complex multiscale systems, especially those involving multiphysics processes. Furthermore, we illustrate the use of the AD-enhanced FFT method for problems that extend beyond homogenization, such as uncertainty quantification and topology optimization where automatic differentiation simplifies the computation of sensitivities. Our work will simplify the numerical implementation of FFT-based methods for complex solid mechanics problems.
  
\end{abstract}

\begin{keyword}
Spectral method, Automatic differentiation, Multiscale, Homogenization
\end{keyword}

\end{frontmatter}


\section{Introduction}

A crucial aspect of computational multiscale mechanics is understanding how microscale mechanisms impact macroscale material behavior. Features such as the distribution of heterogeneities, the formation of structures at the micro level, and physical processes occurring on the microscale can significantly influence the properties homogenized at a larger scale. The Fast-Fourier Transform (FFT) methods~\cite{moulinec_numerical_1998, michel_computational_2001, vondrejc_fft-based_2014, zeman_finite_2017} are particularly well-suited for tackling such ng homogenization problems due to their computational efficiency~\cite{schneider_review_2021, ladecky_optimal_2023}, ease of implementation~\cite{de_geus_finite_2017}, and low memory footprint. Nonetheless, certain aspects of FFT frameworks limit their quick adaptation to various applications and extension to intricate mechanical systems.

A significant challenge with FFT-based methods is that they still rely on manually implemented stress-strain relationships, which are computed ``by hand'' either analytically or through numerical differentiation~\cite{schneider_review_2021}. This makes the implementation of complex constitutive relations (\textit{e.g.}, hyperelastic materials and elastoplastic materials) intricate and error-prone within the FFT framework, especially for cases involving multi-physical processes~\cite{rothe_automatic_2015}. Hence, one of their major strength (\textit{i.e.} ease of implementation) is lost for more complex problems. Another prominent limitation lies in determining the homogenized tangent stiffness in multiscale simulations~\cite{rambausek_two-scale_2019, gierden_efficient_2021, kochmann_efficient_2018, felder_multiscale_nodate, tran-duc_efficient_2024}. Current FFT-based multiscale frameworks require auxiliary solution schemes~\cite{gierden_review_2022} to couple microscale mechanisms to the macroscale through a constitutive tangent stiffness, which can be computationally expensive and inaccurate for complex material models and large computational domains. 

Other numerical methods have similar challenges, which have been overcome by adopting a differentiable framework, which automates the process of computing derivatives~\cite{margossian_review_2019}. This allows the formulation of the governing equation in a high-level manner (usually as an expression of the functional energy), and the complex Jacobian and the Hessian are then strictly derived through automatic differentiation (AD) up to machine precision~\cite{rothe_automatic_2015, vigliotti_automatic_2021}. Among prominent examples of such differentiable frameworks within the field of solid mechanics is FEniCSx~\cite{baratta_dolfinx_2023} for Finite Element Methods. Concurrently, with the rise of AD libraries such as PyTorch~\cite{ansel_pytorch_2024} and JAX~\cite{bradbury_jax_2018}, the differentiable framework has become quite appealing, which is evident by the shift in porting existing frameworks such as molecular dynamics, discrete element method, and finite element method~\cite{schoenholz_jax_2020, carrer_learning_2024, toshev_jax-sph_2024, xue_jax-fem_2023, bleyer_numerical_2024} to a differentiation-based framework. Similarly, automatic differentiation presents great potential for FFT-based methods~\cite{bluhdorn_automat_2022} but has yet to be developed in a general and ease-of-use manner to overcome the main challenges of FFT-based methods.

In this paper, we enhance the FFT framework by incorporating automatic differentiation. We investigate the capabilities of an AD-enhanced FFT framework and examine how automatic differentiation mitigates some of the aforementioned issues. Thereby making FFT-based frameworks more modular and easy to implement for complex solid mechanics problems.

\section{How automatic differentiation works?}
\label{sec:automatic-differentiation}

Consider a function $f : \mathbb{R}^n \to \mathbb{R}^m$ that maps an $n$-dimensional input $x$ to an $m$-dimensional output. To determine the derivatives of the outputs relative to the inputs, specifically $J_{ij} = \partial f_i / \partial x_j$ where $i \in m$ and $j \in n$, one typically depends on the hand-calculated closed-form expression of the derivative of the function. In case the function $f$ has a complex closed-form mathematical expression, one can use symbolic differentiation to get the derivatives~\cite{inc_mathematica_nodate}. In computational solid mechanics, $f$ may represent a complex function consisting of numerous procedures that do not possess an explicit mathematical expression. For example, $f$ might be a function that computes the Fourier transform of the input data with a conditional statement (if-then-else) or operates a minimization solver with for-loops on the inputs. In such cases, neither analytical nor symbolic differentiation is feasible, requiring the use of alternative approaches to estimate the derivatives $\partial f_i/\partial x_j$. A classical approach is numerical differentiation. Its results are very sensitive to the perturbations used in the calculations, which may result in inaccuracies or numerical stability issues. Alternatively, automatic differentiation overcomes such limitations by computing exact derivatives for such functions. This is achieved by decomposing a complex function $f$ into elementary functions, for which exact derivatives are known, and interconnecting them through basic arithmetic operations such as addition and subtraction. The chain rule of calculus is then applied to compute exact partial derivatives. For instance, if the function $f$ can be decomposed as $f = h(g(x))$, such that $x \in \mathbb{R}^n, g: \mathbb{R}^{n} \to \mathbb{R}^k$ and $h: \mathbb{R}^{k} \to \mathbb{R}^m$, then the partial derivative  $df_i/dx_j$ is computed as 
\begin{align}\label{eq:chain-rule}
J_{ij} = \dfrac{\partial f_i}{\partial x_j} =\sum_{p=1}^{k} \dfrac{\partial h_i}{\partial g_p}\dfrac{\partial g_p}{ \partial x_j}~.
\end{align}
As automatic differentiation utilizes the exact derivatives of the sub-functions to determine the derivative of $f$, it is not affected by issues related to numerical stability or round-off errors. The derived values retain the same level of precision as the function results, thereby enabling the accurate calculation of derivatives even in the case of nonlinear or non-monotonic functions. Moreover, automatic differentiation can be repeatedly applied to a function to derive exact higher-order derivatives without accumulating errors. 

\begin{figure}[t]
    \centering
    \includegraphics[width=0.95\textwidth]{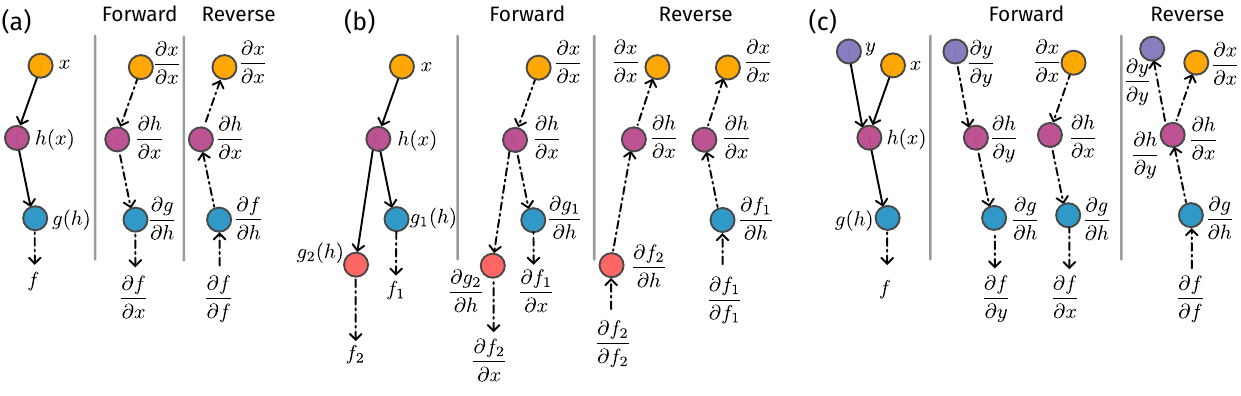}
    \caption{Forward vs. Reverse mode for a composable function $f : g \circ h$ (a) Schematic illustration showing automatic differentiation applied to function  $f : \mathbb{R} \to \mathbb{R}$. The left column shows the execution graph for computing output of $f$. The center column illustrates the execution graph for calculating the derivative of $f$ via forward-mode. Observe that each node representing a function call is substituted by its derivative and is traversed in the original function call direction. The right column indicates the call direction for the reverse-mode, note the backwards arrows. (b)  Schematic illustration showing automatic differentiation applied to function  $f : \mathbb{R}^{1} \to \mathbb{R}^{2}$.  Notice that the forward-mode differentiation requires only a single pass to compute the 2 partial derivatives whereas reverse-mode requires 2 passes to compute the partial derivatives. The 2 passes are indicated by the 2 separate graphs. (c) Schematic illustration showing automatic differentiation applied to function  $f : \mathbb{R}^{2} \to \mathbb{R}^{1}$. Here the forward-mode requires 2 passes to compute the partial derivatives with respect 2 input values whereas reverse-mode requires a single pass.
 }
     \label{fig:fwd-rev}
\end{figure}

Automatic differentiation has two ways to calculate the derivative of $f$ using the chain rule~\cite{margossian_review_2019, bradbury_jax_2018}. The first technique involves evaluating the derivatives of sub-functions (inner derivatives) simultaneously with the function evaluation. This approach is known as \textit{forward-mode} differentiation. \Cref{fig:fwd-rev}a illustrates forward-mode differentiation applied to a composite function $f=g\circ h$, mapping a scalar to a scalar, which requires only one pass to compute both the output of the function and its gradient. Alternatively, the second method involves first evaluating the function's output and then moving backwards to determine the derivatives, starting with outer derivatives followed by inner ones. This is called \textit{reverse-mode} differentiation and requires at least two evaluations to extract the function's value and gradient. \Cref{fig:fwd-rev}a depicts this backward propagation of the function to first compute outer derivatives and then inner ones. The choice of method depends on how the function $f$ maps the input space to the output space. We demonstrate this with two simple cases for a function $f : \mathbb{R}^{n} \to \mathbb{R}^m$: (i) when $n < m$, \ie{}, $f=(g_1(h(x)), g_2(h(x)))$ as shown in \Cref{fig:fwd-rev}b, and (ii) when $n > m$, i.e., $f=g(h(x, y))$ as shown in \Cref{fig:fwd-rev}c. In the case of $n=1 < 2=m$ (see \Cref{fig:fwd-rev}b), forward-mode requires a single pass to calculate both partial derivatives ($\partial f_1/\partial x, \partial f_2/\partial x$), while reverse-mode needs two backward computations to determine these derivatives. Conversely, when $n=2 > 1=m$, forward-mode requires two evaluations: one using $x$ (assuming $y=0$) and another using $y$ (assuming $x=0$), for computing partial derivatives ($\partial f/\partial x, \partial f/\partial y$). However, reverse-mode achieves this with a single pass. Therefore, as a basic rule, for a function $f : \mathbb{R}^{n} \to \mathbb{R}^{m}$, forward-mode is computationally efficient when mapping from a lower-dimensional space to a higher-dimensional one, \ie{}, when $n \ll m$. In contrast, reverse-mode is efficient when the function maps from a higher to a lower dimension space, indicating $n \gg m$. We use this guideline throughout the paper to find the most efficient way to differentiate.

For this study, we use JAX for automatic differentiation due to its effective support for both forward- and reverse-mode differentiation, just-in-time (JIT) compilation, and efficient parallelization on GPUs. JAX is particularly well suited for scientific computing tasks requiring high-order derivatives, offering both flexibility and high performance. 

\section{Differentiable FFT-framework for solid mechanics}
\label{sec:method}

For the purpose of this work, we incorporate automatic differentiation within the Fourier-Galerkin approach~\cite{vondrejc_fft-based_2014} for FFT-based solvers. To this end, we explore the specific arithmetic operations made more straightforward by automatic differentiation and present the relevant pseudo-code/algorithm.

Consider a Representative Volume Element (RVE) $\rve$ subjected to an overall strain $\overline{\strain}$. The governing equation for static mechanical equilibrium in this RVE is thus given by $\nabla\cdot{}\stress = \vector{0} $, where $\stress$ is a function/operator that maps the overall deformation to the appropriate local stress field. Under small-strain conditions, the overall deformation $\overline{\strain}$ represents small-strain $\smallstrain$, and $\stress$ represents a function that maps local $\smallstrain$, to the Cauchy stress tensor $\cauchy$. Conversely, for finite deformations, $\overline{\strain}$ represents the deformation gradient $\overline{\deformation}$ and $\stress$ represents a function that maps the local deformation gradients $\deformation$ to the first Piola-Kirchhoff stress tensor $\pkI$. Since this paper addresses both conditions, we will use the stress operator $\stress$ as a generic term for stresses in the description of the method throughout this section. We note that, consequently, the derivations and procedures remain consistent for both assumptions.

In this study, we adopt the Fourier-Galerkin spectral method~\cite{vondrejc_fft-based_2014, zeman_finite_2017}, and do so for several reasons. Primarily, the variational structure of this approach retains the convexity of the energy density~\cite{zeman_finite_2017, schneider_review_2021}, which ensures that a unique solution to a given problem exists. Secondly, to guarantee that the solution produces a compatible strain state, the method employs a projection operator $\mathbb{G}$, which converts an arbitrary strain state to a compatible state (periodic and curl-free). This operator is independent of the material model and remains consistent for both small-strain and finite deformation scenarios~\cite{de_geus_finite_2017}. %
We will show that these characteristics of the Fourier-Galerkin approach make integrating automatic differentiation within the FFT-based framework straightforward. 

Applying the Fourier-Galerkin method, we reformulate the static equilibrium equation as 
\begin{align}\label{eq:fft-form}
\ifft{ \fft{\mathbb{G}} :   \fft{\stress}} = \vector{0}~,
\end{align}
where $\wavevector$ is the wavevector, $\fft{\star}$ is the Fourier transform of quantity $\star$, and $\ifft{\star}$ is the inverse Fourier transform. For details on the derivation of the above equation and the exact form of the projection operator $\mathbb{G}$, please refer to~\cite{  vondrejc_fft-based_2014, zeman_finite_2017,  de_geus_finite_2017, pundir_fft-based_2023}.  
The above equation is solved using a Newton-Krylov solver, where the Newton-Raphson loop is employed to handle nonlinearities, and a Krylov-subspace solver, such as Conjugate Gradient, is employed to enforce the compatibility conditions (for details, see~\ref{app:newton-krylov}).

Here, we will show how automatic differentiation can be applied in an FFT-based framework. We will describe three use cases of automatic differentiation, namely the stress computation in elastic and hyperelastic materials, the computation of consistent tangent operator for nonlinear problems, and the computation of homogenized stiffness operator for multiscale simulations. For all cases, we include both the arithmetic operations as well as the counterpart pseudo-code to highlight the general applicability and ease of use of automatic differentiation in FFT-based methods. To this end, we employ the open-source JAX library~\cite{bradbury_jax_2018} for automatic differentiation and demonstrate that only a few lines of code are needed to implement a general solution.

\subsection{Automatic computation of stress tensor from energy functional}

We consider the stress computation of elastic and hyperelastic materials, for which the stress function $\stress$ is the first derivative of the strain energy density, as given by
\begin{align}\label{eq:compute-stresses}
\stress(\vector{x}) = \dfrac{\partial \psi(\strain)}{\partial \strain}~.
\end{align}
The application of automatic differentiation on Eq.~\ref{eq:compute-stresses} using the JAX library~\cite{bradbury_jax_2018} is straightforward as we show in \Cref{code:rev-mode}.
\begin{listing}
\begin{minted}[frame=lines,
framesep=1mm,
baselinestretch=0.7,
fontsize=\footnotesize,
]{python}
def strain_energy(strain):
    # compute the energy based on the input `strain`
    return energy

# create a function to compute stresses using the
# derivative of energy  w.r.t `strain` (Eq 2)
compute_stresses = jax.jacrev(strain_energy)

# using the above created function to compute stresses
sigma = compute_stresses(strain)
\end{minted}
\caption{Automatic differentiation to compute stress from energy density functional}
\label{code:rev-mode}
\end{listing}
Here, we define a function that receives microscale strain as input and outputs the corresponding energy density. Since the function maps a $2^\mathrm{nd}$-order strain tensor to a scalar value, we use the reverse-mode technique to differentiate the function to calculate stresses. We note that the input strain to the strain-energy function may be either a small strain ($\smallstrain$) or a deformation gradient ($\deformation$) based on the defined energy density. In Sec.~\ref{sec:energy-derivation}, we will present examples for various material models under small-strain and finite deformation settings.

\subsection{Automatic computation of tangent moduli and incremental stresses}\label{sec:tangent-modulii}

We integrate automatic differentiation to ease the numerical implementation of the nonlinear \Cref{eq:fft-form}. In standard practice, \Cref{eq:fft-form} is linearized, and the Newton-Raphson method is used to find the equilibrium. The linearization process involves computing the incremental stresses and the consistent tangent operator at each step of the Newton-Raphson scheme, which is expressed as follows,
\begin{align}\label{eq:incremental-stress}
 \delta \stress = \underbrace{\dfrac{\partial \stress(\strain)}{\partial \strain}}_{\mathbb{K}}\delta \strain~.
\end{align}
Consequently, the local tangent stiffness $\mathbb{K}$ is needed at the previous strain to calculate the incremental stresses. With automatic differentiation, there are three different ways in which one can differentiate the stress function to compute the local tangent stiffness. The straight-forward approach will be to simply differentiate the obtained stress function again. Since the stress function maps $2^\mathrm{nd}$-order strain tensor to $2^{\mathrm{nd}}$-order stress tensor, one can choose either the forward-mode or the reverse-mode for differentiation. Both will be computationally equivalent. The code-snippet for computing tangent stiffness using reverse-mode differentiation is given in \Cref{code:tangent-rev-mode}. This method explicitly computes and stores the local tangent stiffness in a matrix form which then must be multiplied with the incremental strains $\delta\strain$ to get the incremental stresses $\delta\stress$. This results in extra memory usage and two operations. 
\begin{listing}
\begin{minted}[frame=lines,
framesep=1mm,
baselinestretch=0.7,
fontsize=\footnotesize,
]{python}
# create a function to compute stresses
compute_stresses = jax.jacrev(strain_energy)

# computing the function to compute local tangent modulii K as given in Eq 4
compute_local_tangent = jax.jacfwd(compute_stresses)

# compute the tangent modulii
tangent_modulii = compute_local_tangent(strain)

# compute the incremental stresses as given in Eq 4
dsigma = jax.numpy.einsum('ijkl, kl -> ij', tangent_modulii, dstrain)
\end{minted}
\caption{Automatic differentiation to compute incremental stresses using forward-mode AD}
\label{code:tangent-rev-mode}
\end{listing}

Automatic differentiation and especially the JAX library provides two other methods, \ie{} the jvp and linearize functions, to ease the calculation of directional derivatives. In both approaches, the local tangent stiffness can be represented as a push-forward function (or a linear operator) that maps incremental strains $\delta \strain$ lying in the tangent space of $\strain$ to incremental stresses $\delta \stress$ lying in the tangent space of the stress domain. This facilitates the direct computation of incremental stress as a Jacobian-vector product~\cite{hirsch_differential_2013, balestriero_fast_2021}. 
The jvp function (an example code-snippet is shown in \Cref{code:jvp}) internally constructs the push-forward map and then applies it to the incremental strains to compute the incremental stresses. Representing $\mathbb{K}$ as a push-forward map rather than storing it explicitly in matrix form can significantly reduce the memory usage of FFT-based methods. However, creating this push-forward map with each function call can prove computationally expensive, particularly when utilized in the inner solver, like the conjugate gradient, of a Newton-Krylov solver. 
\begin{listing}
\begin{minted}[frame=lines,
framesep=1mm,
baselinestretch=0.7,
fontsize=\footnotesize,
]{python}
# create a function to compute stresses
compute_stresses = jax.jacrev(strain_energy)

# computing the incremental stresses as given in Eq 4
dsigma = jax.jvp(compute_stresses, (prev_strain,), (dstrain,))[1] 
\end{minted}
\caption{Automatic differentiation to compute incremental stresses using jvp function}
\label{code:jvp}
\end{listing}

A computationally more efficient approach involves computing the local tangent stiffness around a fixed point, such as $\strain$ from the previous converged step, within the Newton-Raphson scheme. This local stiffness can then be used to calculate incremental stresses during iterations in the inner solver. We utilize the JAX linearize function to generate the push-forward map for $\mathbb{K}$ at a specified strain point and store it as a function, which is then utilized to compute the incremental stresses. An example implementation is given in \Cref{code:linearize}.
\begin{listing}
\begin{minted}[frame=lines,
framesep=1mm,
baselinestretch=0.7,
fontsize=\footnotesize,
]{python}
# create a function to compute stresses
compute_stresses = jax.jacrev(strain_energy)

# compute the tangent stiffness function linearized around previous strain
f_jvp = jax.linearize(compute_stresses, prev_strain)[1]

# compute the incremental stresses as given in Eq 4
dsigma = f_jvp(dstrain)
\end{minted}
\caption{Automatic differentiation to compute incremental stresses using linearize function}
\label{code:linearize}
\end{listing}

We examine the computational efficiency of these three methods to determine the incremental stresses from tangent stiffness on two scenarios: calculating incremental stresses for a single pixel (see \Cref{fig:method-efficiency}a) and for a grid of pixels (see \Cref{fig:method-efficiency}b). We consider a linear elastic material law for which the strain energy is given as $\psi = \lambda \trace{\smallstrain}^2 + \mu \trace{\smallstrain:\smallstrain}$ where $\lambda$ and $\mu$ are Lame's parameters. For reference, we show the computational time when the local stiffness tangent and the incremental stresses are computed from the analytical expression. We observe that the calculation time for a pixel in forward or reverse mode, which are the two options for the first approach, is nearly as fast as using the analytical method (see \Cref{fig:method-efficiency}a). A small overhead occurs when using automatic differentiation due to the combined computation of derivatives and function values. However, employing the push-forward map via the linearize function or the jvp function increases the overall computation time for a single pixel (see \Cref{fig:method-efficiency}a). Although direct evidence for this cost increase is unavailable, we speculate that it results from constructing the push-forward map. When considering the computation time of the incremental stresses on a pixel grid (\textit{e.g.} $63^2$, $127^2$, $255^2$), we observe that all the tested methods are on par with the analytic approach (see \Cref{fig:method-efficiency}b). Our results show that the jvp and linearize functions are slightly faster than the forward or reverse mode differentiation. 

\begin{figure}[t]
    \centering
    \includegraphics[width=0.95\textwidth]{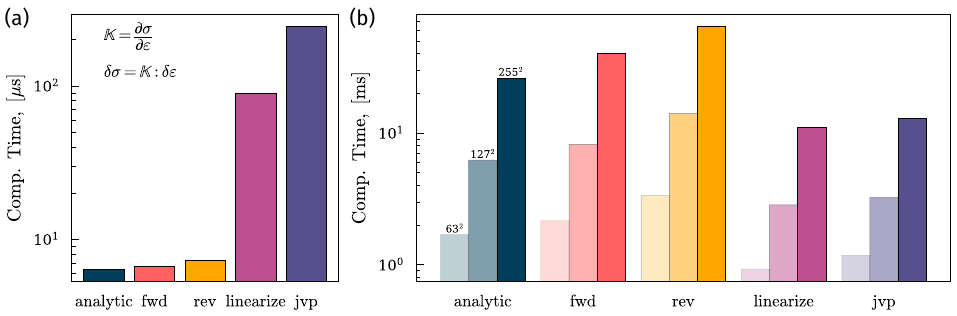}
    \caption{Computational efficiency of AD methods for the computation of incremental stresses from tangent stiffness. (a) Computational time for calculation on a single grid point. We consider a single function evaluation.  (b) Computational time for a varying number of grid points. We compute the incremental stresses simultaneously overall the grid points.
 }
     \label{fig:method-efficiency}
\end{figure}

\subsection{Automatic computation of homogenized stiffness tensor}

We exploit the automatic differentiation technique to compute the homogenized stiffness tensor, which is the most significant aspect of computational homogenization~\cite{eshelby_determination_1997,  blanc_homogenization_2023} or the essential requirement for performing multiscale simulations~\cite{gierden_review_2022, geers_multi-scale_2010}. The effective stiffness tensor is the partial derivative of macroscale stresses with respect to macroscale strains, which reads
\begin{equation}
    \overline{\mathbb{C}} = \dfrac{\partial \overline{\stress}}{\partial \overline{\strain}}, \quad \text{where} \quad \overline{\stress} = \dfrac{\int_{\rve} \stress( \strain(\vector{x})) d\rve }{\int_{\rve} d\rve}~,
\end{equation}
where $\overline{\stress}$ is the volume average of local stresses within the RVE. In the FFT-based framework, two methods are used to determine the effective properties from the local stress and strain distribution: (a) the perturbation approach~\cite{ gierden_review_2022, meier_determination_2016} and (b) the algorithmic consistent approach~\cite{gokuzum_algorithmically_2018}. In the perturbation approach, each component of the effective tensor is computed by numerical differentiation (finite difference) between the perturbed state and the reference state. In the algorithmic consistent approach, the consistent tangent operator is computed by solving the Lippmann-Schwinger equation~\cite{gokuzum_algorithmically_2018}. Both approaches necessitate an auxiliary solution method alongside the primary one to determine effective properties. For instance, the perturbation method requires computing finite differences for every tensor component, leading to three computations in a 2D case and six in a 3D case. Similarly, resolving the Lippmann-Schwinger equation requires an iterative solver. Consequently, both approaches can be computationally intensive, especially for large computational domains or for nonlinear constitutive laws. In contrast to these approaches, automatic differentiation significantly simplifies the procedure by employing a single operation for the computation of the homogenized tangent stiffness, as shown in \Cref{code:macro-tangent}.
\begin{listing}
\begin{minted}[frame=lines,
framesep=1mm,
baselinestretch=0.7,
fontsize=\footnotesize,
]{python}
# function to solve RVE  using FFT approach for a given macroscale strain
def solve_microscale(macro_strain):
    # solve RVE (Eq 1) using Newton-Krylov solver as given in Appendix A 
    # and get the microscale strain
    local_strain = netwon_krylov_solver(macro_strain)

    # compute the microscale stresses from microscale strain
    local_sigma = compute_stresses(local_strain)

    # compute the macroscale stresses from microscale stress
    macro_sigma = mean(local_sigma)
    return macro_sigma, (macro_sigma)

# create a function to compute homogenized tangent and stresses
compute_macro_tangent_stiffness = jax.jacfwd(
    solve_microscale, argnums=0, has_aux=True
)

# calling tangent function to get macroscale stresses and tangents
macro_tangent, macro_stress = compute_macro_tangent_stiffness(macro_strain)
\end{minted}
\caption{Automatic derivation of homogenized stiffness tensor}
\label{code:macro-tangent}
\end{listing}

Here, we directly differentiate a function that takes macro-strain as input and outputs macro-stress (after solving \Cref{eq:compute-stresses}) to compute the homogenized stiffness tensor. This function encompasses numerous complex operations or function calls, such as forward-FFT transform, inverse FFT transform, and a Newton-Krylov solver, to accurately determine the local strain and stress distribution in an RVE (for details, refer~\cite{de_geus_finite_2017, pundir_fft-based_2023}). The ability of automatic differentiation to decompose complicated arithmetic operations into primitive ones allows for exact differentiation. Moreover, it facilitates the simultaneous computation of stresses and tangent stiffness, thus eliminating the need for any secondary solution method. 

\section{Applications}
\label{sec:applicatons}

Incorporating automatic differentiation in an FFT-based framework can, as described in \Cref{sec:method}, greatly simplify the computation of complex derivatives, which is essential for many mechanical problems. Here, we employ the AD-enhanced Fourier-Galerkin method on problems of varying complexity to demonstrate its capabilities. Specifically, we examine mechanical problems with intricate material models, bodies undergoing 
large-strain deformation, and multiscale simulations.  We show how automatic differentiation allows easy differentiation of functions with complex mathematical operations such as FFT transform or with logical operations such as conditional/unconditional-loops, which simplifies the entire numerical framework. We evaluate the computational performance and accuracy of the AD-enhanced FFT-based method.   We provide the Python code for all the examples at \cite{pundir_supplementary_2024}. All simulations were performed as a serial job on a Linux system using a 4-core AMD EPYC™ 9654 CPU with 16 GB RAM and a clock speed of 2.4Hz.

\subsection{Automatic derivation of stress and tangent operator}
\label{sec:energy-derivation}

We present two examples to demonstrate the applicability of automatic differentiation for the computation of stress and tangent operators for material laws with fundamentally different characteristics, \ie{} hyperelastic and elastoplastic materials.

\begin{figure}[t]
    \centering
    \includegraphics[width=0.95\textwidth]{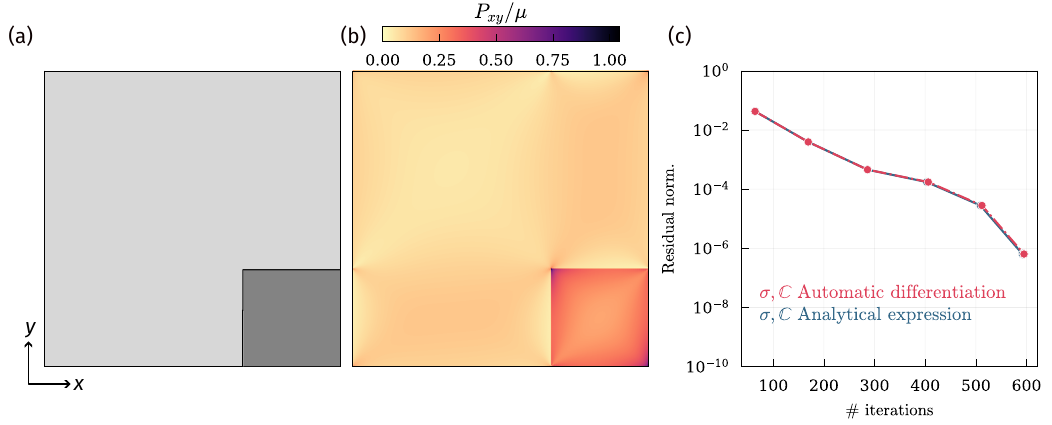}
    \caption{Automatic computation of stress and tangent operator for a St.~Venant-Kirchhoff material. (a) Schematic illustration of the microstructure of the considered RVE.  (b) Local distribution of the first Piola-Kirchhoff stresses within the RVE. (c) Variation of the residual norm and the cumulative CG iterations for each NR iteration (shown as circles). 
    }
     \label{fig:st-venant}
\end{figure}

\subsubsection{Example on hyperelastic material}\label{sec:hyperelastic}
With this example, we aim to show that the reliance of FFT-based frameworks on explicit formulations for stress and tangent operators can be minimized~\cite{schneider_review_2021}. We consider \textit{hyperelastic} materials, where the stresses and the tangent moduli are first- and second-order derivatives of the strain energy density functional, respectively. We consider the St.~Venant-Kirchhoff material for which the energy density functional is given as
\begin{align}
\psi &= \dfrac{\lame}{2}\trace{\greenstrain}^2 + \shearmodulus \trace{\greenstrain:\greenstrain}~,
\label{eq:hyperelastic}
\end{align}
where $\greenstrain = \deformation\cdot{}\deformation^\mathrm{T} - \tensor{I}$ is the Green-Lagrange strain, $\deformation$ is the deformation gradient, $\lame$ and  $\shearmodulus$ are Lame's parameters. As described in \Cref{sec:method}, we apply automatic differentiation to derive the stress function (the Jacobian of the energy density) and the tangent function (the Hessian of the energy density). 

Foremost, we compare the AD implementation for St.~Venant-Kirchhoff material with the implementation where the stress and tangent functions are computed from the analytical expression. Similar to~\cite{de_geus_finite_2017}, we consider a square RVE of dimension $\ell$ (see \Cref{fig:st-venant}a) with a square stiff inclusion (of dimension $\ell/3$) at one of its corners. The RVE is subjected to a macroscopic shear loading in terms of  $\macro{\deformation}$ which in Voigt notation is given as $ [1, 1,  1+0.1\ell]$. We discretize the RVE into 299$^2$ grid points. An odd number of grid points is chosen to maintain the compatibility of the deformation gradient~\cite{zeman_finite_2017}. We solve the problem in one increment using a Newton-Krylov solver with a conjugate gradient (CG) solver for solving the linearized form of \Cref{eq:fft-form}. The tolerance for the outer  Newton-Raphson (NR) loop is $10^{-5}$, and the tolerance for the inner loop of the CG solver is $10^{-8}$. We compute the local distribution of the first-Piola Kirchhoff stresses within the RVE (see \Cref{fig:st-venant}b) and the residual norm ($\delta \deformation / \| \macro{\deformation} \|$) for every NR iteration (see \Cref{fig:st-venant}c). Here, we use JAX's jvp function (see \Cref{sec:tangent-modulii}) to compute the incremental stresses within the CG iterations. The AD implementation gives identical results to the analytical expression, as the residual norm and the iteration counts (cumulative CG iterations for every NR iteration) to reach convergence are the same. The computational times for the simulations with AD implementation and with analytical implementation are approximately the same, around 19.5 seconds. Since we express the tangent operator as a function rather than storing it as a matrix, the memory footprint of the AD-enhanced version of our implementation is roughly 1/2 of that of the analytical expression. 

This example demonstrates that stress and the tangent operator can be easily derived from an energy functional for a material. This approach eliminates the need for FFT-based methods to rely on analytical expressions or numerical differentiation, thereby reducing computational cost (\textit{i.e.}, memory usage) and simplifying the numerical implementation of sophisticated hyperelastic materials. 

\begin{figure}[!t]
    \centering
    \includegraphics[width=0.95\textwidth]{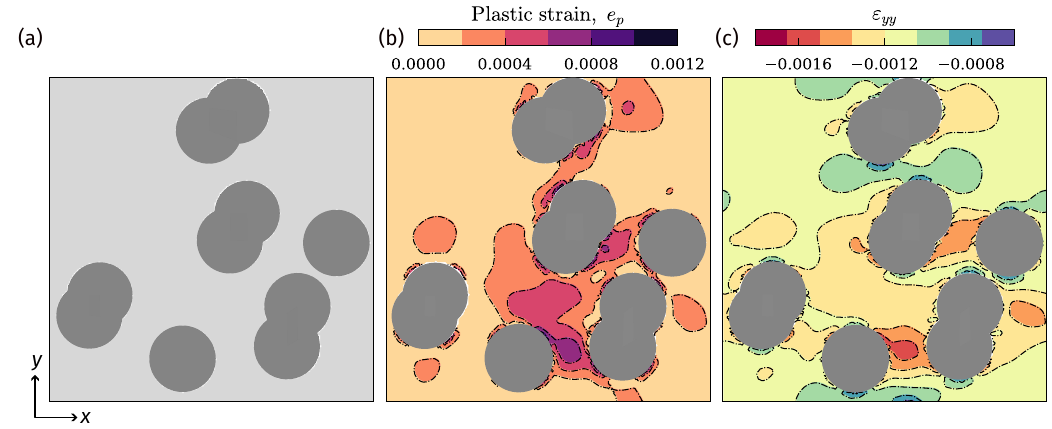}
    \caption{Automatic computation of tangent operator for $J_2$-plasticity. (a) The microstructure of the considered RVE. The hard circular inclusions have a higher yielding strength than the matrix.ribution of cumulative plastic strain $\varepsilon_{p}$ and (c) local distribution of strain within the RVE. The color map is the result of AD-enhanced FFT simulations, and the dashed contour lines represent the results of the analytical implementation of the elastoplastic material.}
    \label{fig:elastoplastic}
\end{figure}

\subsubsection{Example on elastoplastic material}\label{sec:elastoplastic}
Automatic differentiation is applicable to material laws in which the stresses and the tangent stiffness cannot be explicitly derived from the energy density. For instance, elastoplastic materials involve stress and tangent computations that depend on yielding criteria, plastic flow rule, and a return-mapping algorithm~\cite{simo_computational_1998}. For demonstration purposes, we consider a heterogeneous RVE with hard circular inclusions embedded in a soft matrix (see \Cref{fig:elastoplastic}a). We adopt an isotropic linear hardening model in a small-strain setting with each phase exhibiting distinct hardening exponents and yield stresses. The RVE is subjected to a macroscopic biaxial loading, in terms of $\macro{\smallstrain}$ which in Voigt notation is given as $[10^{-2}\ell, 10^{-2}\ell, 0]$. The RVE is discretized into $199^{2}$ grid points. The tangent stiffness operator in elastoplastic materials depends on the elastic strain, the plastic strain, the type of yield surface, and the normals to this yield surface~\cite{belytschko_nonlinear_2014, simo_computational_1998}.  We selected $J_2$-plasticity for both materials.  We assume a linear isotropic hardening law with the yield criterion given as $\Phi = \sqrt{3/2 \stress_ \text{dev}:\stress_\text{dev}} - \sigma_y$ where $\stress_\text{dev}$ is the deviatoric stress and $ \sigma_y$ is the yielding stress. The phase contrast between the yield stress and hardening moduli of the two phases is 2. The analytical expression for the consistent constitutive tangent is equal to the elastic tangent when the trial state is elastic and otherwise given as
\begin{align*}
\mathbb{K} &= \frac{\partial \stress^{t+\Delta t}}{\partial \strain^{t + \Delta t}} = \mathbb{C}_\text{elas} - \frac{6\mu^2 \Delta \epsilon_p}{\text{tr} \sigma_{\text{eq}}} I_d + 4\mu^2 \left( \frac{\Delta \epsilon_p}{\text{tr} \sigma_{\text{eq}}} - \frac{1}{3 \mu + H} \right)     \text{tr} N \otimes \text{tr} N ~.
\end{align*}
where $\mathbb{C}_\text{elas}$ is the elastic stiffness matrix, $\mu$ is the shear modulus, $\epsilon_p$ is the cumulative plastic strain, $H$ is the hardening modulus, $\sigma_y$ is the yielding stress and $N$ is the direction of the deviatoric strain. Automatic differentiation calculates the tangent stiffness on the fly by differentiating the stresses with respect to the strains. Unlike the previous case of hyper-elastic material, where the local stiffness tangent operator is computed as a double derivative of an energy density function, here we first implement a Python function to compute the correct stress state.  The stress state is calculated using the elastic predictor-corrector approach (for details, see \ref{app:elasto-plastic-material}), and a return mapping algorithm~\cite{belytschko_nonlinear_2014, simo_computational_1998}. Then, this function is automatically differentiated with respect to the input. Here, within the NR loop, we use JAX's linearize function (see \Cref{sec:method}) to create the push-forward map of tangent stiffness at a fixed strain value and then use it to calculate incremental stresses within the CG iterations. For details on the numerical implementation of an elastoplastic material for an AD-enhanced FFT-based method, see the pseudo-code in \ref{app:elasto-plastic-material}.

We observe that the local distribution of the plastic strain and the elastic strain (see \Cref{fig:elastoplastic}b and \Cref{fig:elastoplastic}c) for the AD-computed local tangent stiffness operator $\mathbb{K}$ matches the strains computed with the analytical expression of $\mathbb{K}$ (shown with dashed contour lines in \Cref{fig:elastoplastic}). We note that the computation time for the AD implementation and the analytical implementation was approximately the same, around 0.25 seconds. Through this example, we show that a local stiffness operator can also be achieved for the cases where the stresses are not a direct derivative of the energy functional. This increases the applicability of automatic differentiation to various other material models where the calculation of stresses involves complex computational functions or operations.

\begin{figure}[t]
    \centering
    \includegraphics[width=0.95\textwidth]{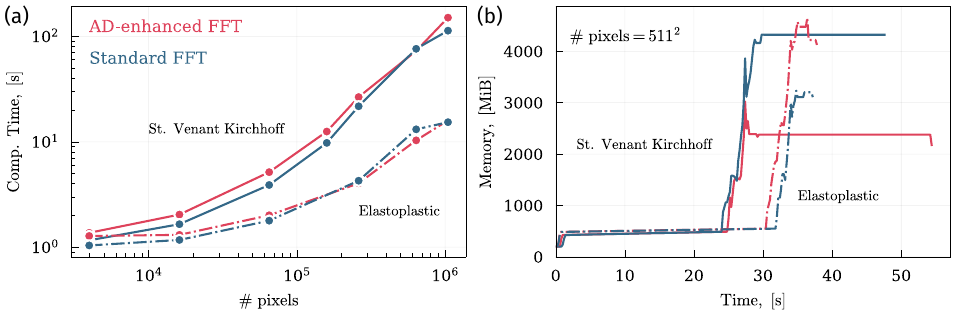}
    \caption{Computational efficiency of AD-enhanced FFT method. (a) Scaling of computational time with increase in the number of grid points for two different constitutive laws. Computations with automatic differentiation and analytical expression are shown in red and blue, respectively. For hyperelastic material, we use JAX's jvp function to compute the incremental stresses and for elastoplastic material, we use JAX's linearize function. (b) Memory consumption for the two material laws for the case with 511$^2$ grid points.
 }
     \label{fig:computational-efficiency}
\end{figure}

\subsubsection{Computational performance}\label{sec:computational-performance}
We present scalability and memory usage (see \Cref{fig:computational-efficiency}) for the two cases considered above, \ie{} hyperelastic and elastoplastic. We report the total computational time required for all stages, starting with the tangent stiffness function derivation and proceeding through Newton-Krylov iterations, up to convergence.  For the hyperelastic case, we applied the push-forward map, utilizing the jvp function, to calculate incremental stresses within each iteration of the Conjugate Gradient solver. As discussed in \Cref{sec:tangent-modulii}, constructing the push-forward map in an inner loop incurs overhead, making the AD-enhanced FFT slightly more expensive than the conventional FFT implementation with analytic expressions. Our scaling analysis shows that even with a high number of pixels, the difference in computation time remains minimal (see \Cref{fig:computational-efficiency}a). However, memory consumption is reduced by approximately half since the tangent moduli are not stored (see \Cref{fig:computational-efficiency}b).

In the elastoplastic case, we computed and stored the push-forward map during Newton-Raphson iterations using the linearize function. The computation time here matches the standard method using analytic expressions because automatic differentiation overhead occurs only in the Newton-Raphson steps (see \Cref{fig:computational-efficiency}a). However, unlike before, the AD-enhanced method uses more memory compared to standard FFT, probably due to the need to store the linearized function (see \Cref{fig:computational-efficiency}b).

Finally, we note that while we showed in \Cref{fig:method-efficiency}b that analytic computations cost more than the push-forward map, these analytic computations become cheaper than linearize and jvp functions for repeated executions. This occurs because the analytic expression, implemented in Numpy, is optimized for multiple uses, whereas the slight overhead of using automatic differentiation makes linearize or jvp marginally computationally more expensive.

\subsection{Computational homogenization}
\label{sec:homogenization}

We apply the AD-enhanced FFT-based framework to perform computational homogenization of an RVE.  As noted earlier, current approaches, such as numerical differentiation or the algorithmic consistent approach, require an auxiliary computationally intensive method to compute the homogenized stiffness tensor of an RVE. Here, we present two examples to illustrate the efficacy and efficiency of automatic differentiation in computing effective properties.

\subsubsection{Validation with Eshelby's problem}
First, we consider the Eshelby problem~\cite{eshelby_determination_1997} as a basic example to compare the numerical result against the analytical solution. Similar to \cite{gokuzum_algorithmically_2018}, we model a periodic micro-structure with a stiff circular inclusion embedded centrally in a compliant matrix.  We compare the numerically computed effective elastic tensor  $\macro{\mathbb{C}}$ with Eshelby's solution. We consider a very low volume fraction (0.7$\%$) of the inclusion to match the infinite medium assumption of Eshelby, similar to~\cite{gokuzum_algorithmically_2018}. We consider a linear elastic material law for both the matrix and the inclusion. Thus, the strain energy  is $\psi= \lambda \trace{\smallstrain}^2 + \mu \trace{\smallstrain :\smallstrain }$. The material parameters chosen for the matrix are $\lambda_m= 2$  GPa and $ \mu_m=1$ GPa, and for the stiff inclusion are $\lambda_{i}=10$ GPa and $\mu_{i}=5$ GPa. We compute the homogenized or effective elastic tensor $\macro{\mathbb{C}}$ of the material for a macro-strain of $\macro{\varepsilon}=[10^{-2}, 10^{-2}, 0]$. We observe that the results from the AD implementation are close to the analytical results (see \Cref{tab:eshelby}). We note that for a computational grid of 199$^2$, the total computation time was around 4 sec. 

 \begin{table}[H]
    \centering
    \begin{tabular}{c|c|c|c}
        &  $\macro{\mathbb{C}}_{xxxx}$ & $\macro{\mathbb{C}}_{yyyy}$ &  $\macro{\mathbb{C}}_{xyxy}$    \\
        \hline
     Analytical    &   4.037  &  4.037  & 1.011 \\
     Numerical & 4.035 &  4.035   & 1.010 
    \end{tabular}
    \caption{Components of the effective elastic tensor $\macro{\mathbb{C}}$ as computed numerically by automatic differentiation using FFT solvers compared against Eshelby's solution for a stiff inclusion embedded in a compliant infinite matrix.  }
    \label{tab:eshelby}
\end{table}

\subsubsection{Example on architected materials}
We apply the AD-enhanced computational homogenization method to an RVE (of dimension $\ell$) with intricate microstructure. We chose architected materials as our example. Architected materials are composed of two or more distinct phase materials arranged in a specific pattern that results in exceptional mechanical properties~\cite{balan_p_auxetic_2023, karapiperis_prediction_2023, magrini_control_2024}. We examine a two-phase linear elastic architected material as illustrated in \Cref{fig:architected}a, comprising a compliant phase and a stiff phase, where the stiff phase is arranged in a hexagonal pattern. We vary the phase contrast $\mathcal{X}=\mathrm{compliant/stiff}$ from 1 to $10^{-4}$  where a value of 1 represents a homogeneous material and value $\ll 1$ approximates a lattice metamaterial~\cite{shaikeea_toughness_2022}. We apply a macroscopic compressive strain $\overline{\smallstrain}$ in \textit{x} and \textit{y}-directions under a small-strain setting. The macroscopic strain tensor in Voigt notation is given as $[ -10^{-3}\ell, -10^{-3}\ell, 0]$. For the case where $\mathcal{X}=1$, \ie{} a homogeneous material, the computed macro stiffness values match the analytical solution (see \Cref{fig:architected}c). With increasing phase contrast, however, the macroscopic stiffness reduces, as expected, and the obtained values ($ \macro{\mathbb{C}}_{xxxx}=0.51, \macro{\mathbb{C}}_{yyyy}=0.42$ and $\macro{\mathbb{C}}_{xyxy}=0.059$) saturates to the values that are indicative of a lattice metamaterials. The values obtained are similar to the values ($\macro{\mathbb{C}}_{xxxx}=0.57, \macro{\mathbb{C}}_{yyyy}=0.52$ and $\macro{\mathbb{C}}_{xyxy}=0.15$) obtained from a full-field Finite Element homogenization under affine displacement boundary conditions (for details, please refer to \ref{app:lattice-material}). The properties obtained under periodic boundary conditions are expected to be lower than the properties obtained under affine displacement boundary conditions~\cite{miehe_computational_2002}.

While not depicted in \Cref{fig:architected} we calculated the homogenized properties for a phase contrast of $10^{-10}$ to approximate zero stiffness in the compliant phase to machine precision. The obtained homogenized values remain consistent, illustrating that the AD-enhanced approach is not affected by significant contrast (approaching zero) in phase properties.

\begin{figure}[!t]
    \centering
    \includegraphics[width=0.95\textwidth]{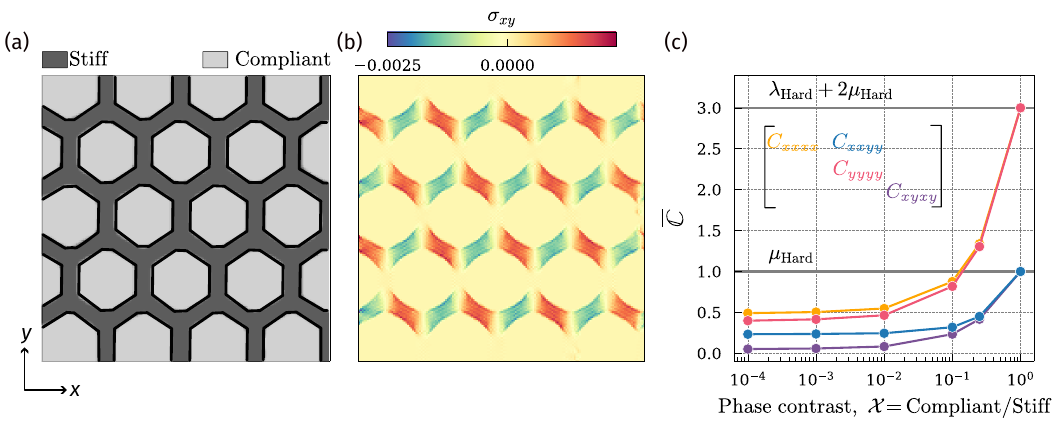}
    \caption{Computational homogenization of an architected lattice material. (a) The architected RVE with a stiff phase arranged in a hexagonal pattern within a compliant phase. (b) Local distribution of shear stresses within the RVE for a phase contrast of $\mathcal{X}=0.5$ and a macroscopic strain $\macro{\strain} = [ -10^{-3}\ell, -10^{-3}\ell, 0]$. (c) Components of the macroscale constitutive tangent stiffness as a function of phase contrast $\mathcal{X}$. The dark grey horizontal lines show the stiffness component for the homogeneous case \ie{} $\mathcal{X}=1$.  }
     \label{fig:architected}
\end{figure}

\subsection{Application to multiscale problem} \label{sec:multiscale}
In this section, we showcase how the ease of computation of homogenized stiffness tensor allows seamless integration of the AD-enhanced FFT-based framework with an additional solver. To this end, we explore a multiscale framework where the macroscale problem is tackled using the Finite Element Method (FEM), while the microscale problem employs the FFT-based approach. This example demonstrates the flexibility of the AD-enhanced FFT-based method by integrating it with a FEM solver. 

The macroscale problem involves a beam with dimensions $L \times H$, where the left end is fixed in both the \textit{x} and \textit{y} directions. A displacement of $10^{-2}L$ is applied to the right edge along the \textit{y}-axis (see \Cref{fig:lattice-multiscale}a). At the microscale, we use the same composite RVE as in \Cref{sec:homogenization} but with the St.~Venant-Kirchhoff material model to capture the geometric nonlinearity (see \Cref{eq:hyperelastic}). A phase contrast of 0.01 is chosen between the compliant and stiff phases to represent voids, thus modeling a hexagonal-lattice metamaterial. We solve the macroscale problem in reference configuration with $\pkI$ and $\deformation$ as the stress-strain measures. The weak form at the macroscale thus reads, 
\begin{align}
    \int_{V} \underbrace{\macro{\mathbb{C}}(\deformation):\macro{\deformation}}_{\macro{\pkI}} : \delta\macro{\deformation}\cdot{}  \dV{V} = 0, \quad \text{where} \quad \macro{\mathbb{C}} = \dfrac{\partial \macro{\pkI}}{\partial \macro{\deformation}}~, 
\end{align}
where the  $\delta\macro{\deformation}$ is the virtual deformation gradient (or test function) and the  $\macro{\mathbb{C}}$ is the homogenized tangent operator computed from microscale. We discretize the beam into 16 quadrilateral finite elements (16 Gauss points) as shown in \Cref{fig:lattice-multiscale}b, where each Gauss point is linked to the RVE with 169$^2$ voxels. The deformation gradient $\macro{\deformation}$ at each Gauss point within the macroscale domain is transferred as input to the corresponding RVE associated with that specific Gauss point. Similar to the example in \Cref{sec:homogenization}, we determine the homogenized tangent operator $\macro{\mathbb{C}}$ through automatic differentiation and pass it together with the average first-Piola Kirchhoff stress $\macro{\pkI}$ to the FEM solver (see \Cref{fig:lattice-multiscale}a). We note that due to a non-symmetric deformation gradient, one needs to use a full deformation gradient (4 components)~\cite{meier_determination_2016}. This means that $\macro{\mathbb{C}}$ is a $4\times4$ matrix.  We perform the simulation in one displacement increment under static conditions and use FEniCSx to manage the FEM component~\cite{baratta_dolfinx_2023, bleyer_numerical_2024}.

\begin{figure}[t]
    \centering
    \includegraphics[width=\textwidth]{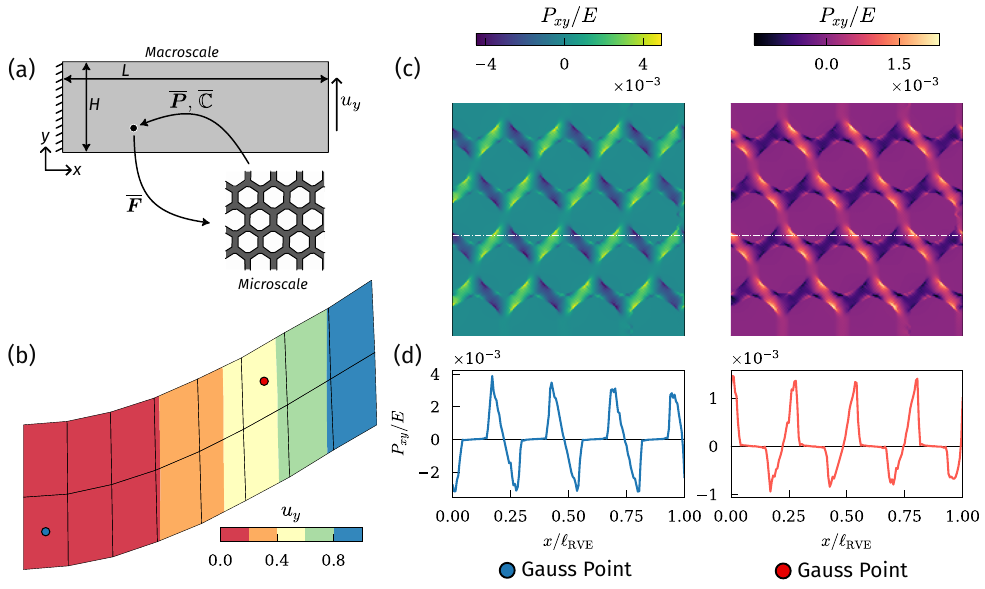}
    \caption{Multiscale simulation of a hyperelastic lattice material. (a) Schematic illustration of the beam at the macroscale and the hexagonal lattice structure at the microscale. (b) The deformed state of the beam for an applied displacement of $10^{-2}L$. The color corresponds to the macroscale displacement along the $y-$axis. The finite element discretization is shown by black lines and two Gauss points are highlighted by colored (red and blue) points. (c) The local shear stress distribution within the RVEs associated with the 2 Gauss points highlighted in (b). (d) Normalized shear stress profile along line highlighted by white dashes in (c).
 }
     \label{fig:lattice-multiscale}
\end{figure}

At the microscale, we apply the Newton-Krylov solver (with an outer tolerance of $10^{-5}$ and an inner tolerance of $10^{-6}$) to accurately capture the nonlinearity within the lattices. At the macroscale, a separate Newton-Raphson solver (with a tolerance of $10^{-8}$) is used to ensure proper convergence. We observe that the chosen contrast of 0.01 between the two phases sufficiently replicates the voids, which is demonstrated by the stresses in the compliant phases dropping to 0, as shown by the stress profile across the cross-section in \Cref{fig:lattice-multiscale}c. We report that this simulation takes approx. 20 minutes in total (for roughly 1 million degrees of freedom in total). A large part of the computational time is spent in the CG-solver because of the high contrast in the phase properties. This is a known issue for the Fourier-Galerkin approach~\cite{zeman_accelerating_2010}, which can be improved using pre-conditioned solvers~\cite{ladecky_optimal_2023}. A traditional FE-FFT framework~\cite{gierden_review_2022}, such as finite-difference-based computation of homogenized stiffness tensor, would require 4 operations per Gauss point, one for each component of non-symmetric deformation gradient, to compute $\macro{\mathbb{C}}$. Conversely, the AD-enhanced multiscale framework is notably advantageous in this scenario, as it demands only one operation per Gauss point for the same.

\subsection{Beyond stress and tangent computations}\label{sec:sensitivity-analysis}
The application of automatic differentiation to compute derivatives extends beyond stress and tangent calculations. Several challenges in solid mechanics require derivatives that are difficult to compute, such as evaluating the sensitivity of a complex computational model to its input parameters. In this section, we illustrate two use cases where automatic differentiation can aid in computing sensitivities of complex computational models such as in uncertainty quantification and topology optimization. Here, we employ the FFT-based method, as presented in earlier sections, as the complex computational model. 

\begin{figure}[t]
    \centering
    \includegraphics[width=0.975\textwidth]{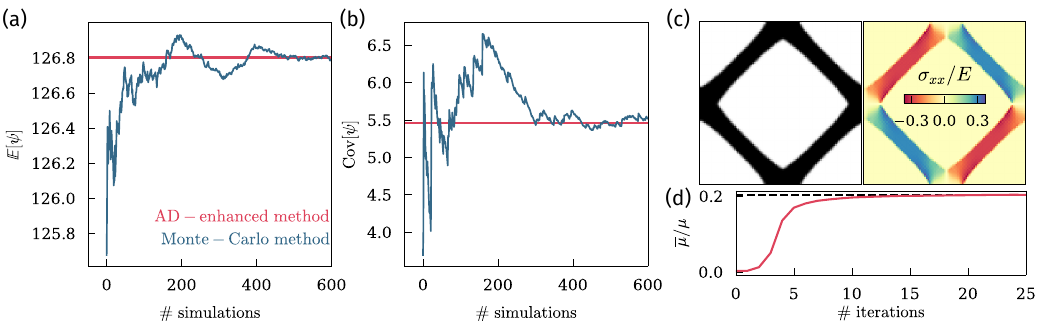}
    \caption{Sensitivity calculation using automatic differentiation for uncertainty quantification and topology optimization (a) Computation of expected value of strain energy for two-phase hyperelastic material using the $2^\mathrm{nd}$-order approximation via AD (red line) and using  the Monte Carlo approach on 600 samples as a reference (blue curve). (b) The covariance in strain energy computed through first-order approximation via AD (red) and through the Monte Carlo approach (blue) on 600 samples as a reference. (c) Microstructure optimized for a maximal homogenized shear modulus at a volume fraction of 30$\%$ (left), and the associated stress distribution (right). (d) The value of the homogenized shear modulus $\overline{\mu}$ as it evolves over optimization iterations. The value is normalized by the shear modulus of the parent material. 
 }
     \label{fig:sensitivity-analysis}
\end{figure}

\subsubsection{Uncertainty Quantification}\label{sec:uncertainity}
We investigate how automatic differentiation can assist in optimizing uncertainty quantification within deterministic processes. Uncertainties may arise from variations in material parameters or inconsistencies in material arrangement. A common method to propagate these uncertainties (\ie{}, calculate the expected and covariance values of the output) employs Monte Carlo simulations, which, while being effective, are computationally demanding. When the input uncertainties are small, the expected value of the output $\mathbb{E}[y]$ and its covariance $\text{Cov}[y]$ can be approximated using the Taylor series expansion. For example, the second-order approximation for the expected value and the first-order approximation for the variance can be expressed as
\begin{align}
 \mathbb{E}[y] \approx f(\mathbb{E}[\vector{x}]) + \dfrac{1}{2} C_{ij}\dfrac{\partial ^2 f(\vector{x})}{\partial x_i \partial x_j} \Big |_{\vector{x}=\mathbb{E}[\vector{x}]}, \quad  \mathrm{Cov}[y]  \approx   C_{ij} \dfrac{\partial f(\vector{x})}{\partial x_i} \Big |_{\vector{x}=\mathbb{E}[\vector{x}]} \dfrac{\partial f(\vector{x})}{\partial x_j}\Big |_{\vector{x}=\mathbb{E}[\vector{x}]},
\end{align}
where $C_{ij}$ is covariance matrix of input parameters, $\mathbb{E}[\vector{x}]$ is the expected value of the input variables $\vector{x}$, $\partial f / \partial x_i$ and $\partial^2 f/\partial x_i \partial x_j$ are the first and the second order derivative of a objective function with respect to inputs $x_i$. Typically, the covariance matrix of the input variables is known and therefore the task is to find the derivatives. Here, we employ automatic differentiation to derive the first derivative and the second derivative of the objective function $f$ with respect to its input. To this end, we consider the same two-phase material as in \Cref{sec:hyperelastic}. The material is \textit{hyperelastic} and the expected values of the material parameters and their variance are given  in \Cref{tab:stat_properties}.
\begin{table}[t]
    \centering
    \begin{tabular}{c|c|c|c|c}

         & $E_{\text{i}}$ [MPa] & $\nu_{\text{i}}$ & $E_{\text{m}}$ [MPa] & $\nu_{\text{m}}$ \\
             \hline
        $\mathbb{E}$ & 5.7 & 0.386 & 0.57 & 0.386 \\
        $\sqrt{\text{Cov}}$ & 0.01 & 0.03 & 0.01 & 0.03 \\
    \end{tabular}
    \caption{The expected and the standard deviation values of the elasticity modulus and the Poisson ratio for the inclusion ($E_\text{i}$ and $\nu_\text{i}$)  and the matrix ($E_\text{m}$ and $\nu_\text{m}$).} 
    \label{tab:stat_properties}
\end{table}

To simplify the analysis, we consider a function $f$ that accepts material parameters as inputs, determines static equilibrium for a macroscopic deformation gradient $\macro{\deformation}= [0,0, 0.1]$, and then outputs the total strain energy of the two-phase material. Given that $f$ maps a 4-dimensional space to a scalar space, we use reverse-mode AD to efficiently obtain the first derivative. We differentiate this newly derived function using forward-mode to compute the second-order derivative. Since the function that evaluates the first-order derivative maps a 4-dimensional space to another 4-dimensional space, forward-mode proves to be efficient. 

We assume that the covariance $C_{ij}$ between the different parameters is known and is computed by drawing 600 samples randomly from a normal distribution. Using the first-order function, the second-order function and the known covariance matrix, we calculate the expected value of the total strain energy and its variance. We note that we only need 2 runs, first for the first-order derivative and second for second-order derivative. The results thus obtained are shown in \Cref{fig:sensitivity-analysis}a~\&~b. For reference, we also determine the expected values and variance through Monte-Carlo simulations using the drawn sample of 600. We observe that the AD-computed values match closely with the Monte-Carlo results (see \Cref{fig:sensitivity-analysis})a~\&~b. Additionally, the total time required for the Monte-Carlo method to achieve a saturated value was approximately 7 minutes, whereas the AD-computed uncertainties took around 20 seconds.

\subsubsection{Microstructure optimization}\label{sec:optimization}

Here, we utilize AD-enhanced FFT-based method for topology optimization and use automatic differentiation to simplify the sensitivity calculation. Topology optimization is frequently utilized to design microstructures with tailored homogenized properties, such as achieving the highest bulk modulus or shear modulus with a specific volume fraction. The microstructure, represented as a density field $\rho(\vector{x})$ varying from 1 (material) to 0 (void), is optimized to achieve the desired properties under specified constraints.  The optimization problem thus reads:

\begin{align*}
    \underset{\rho(\vector{x})}{ \text{minimize}} &:~ f(\rho(\vector{x}), \macro{\smallstrain}) \\
     \text{s.t} &: ~\ifft{ \fft{\mathbb{G}} :   \fft{\stress}} = \vector{0}~, \\
                & : ~\smallint \rho(\vector{x}) d\rve / \smallint d\rve = \vartheta~,  \\
                & :~ 0 \leq \rho(\vector{x}) \leq 1 , \forall \vector{x} \in \rve  
\end{align*}
where $f$ is the objective function to be minimized. The first condition ensures static equilibrium. The second condition ensures the imposed volumetric constraint and the last condition ensures that the density field $\rho(\vector{x})$ remains within physical limits \ie{} $\rho(\vector{x}) \in [0, 1]$. Irrespective of the solution strategy (optimality criteria~\cite{bendsoe_topology_2004}, MMA~\cite{svanberg_method_1987} or Lagrange multipliers) used for the aforementioned optimization problem, information about the gradients of both the objective function and the constraints with respect to the input is required. Recently, automatic differentiation has been adopted to simplify the computation of these gradients~\cite{chandrasekhar_auto_2021}. Building on this concept, here we use automatic differentiation to determine both homogenized properties (\ie{} homogenized shear modulus $\macro{\mu}$) and their sensitivities to the density field within a FFT framework. 

Similar to \cite{xia_design_2015, andreassen_efficient_2011, chandrasekhar_auto_2021}, we consider the classic microstructure optimization problem of maximizing the homogenized shear modulus $\macro{\mu}$ of a microstructure (\ie{} objective function $f(\rho(\vector{x}))=-\macro{\mu}$) with a given volume fraction. We examine a square domain under macroscopic strain $\macro{\smallstrain}$. The material behaves as linear elastic with a Young's modulus $E = 1$ MPa and a Poisson ratio $\nu = 0.3$. Additionally, we enforce a volumetric constraint of 0.3. We follow the optimality criteria method~\cite{bendsoe_topology_2004} to solve the above constrained optimization problem. This method requires the sensitivity of the objective function $\partial f / \partial \rho $ to update the density field $\rho$ in successive iterations. Since the function $f$ maps a high-dimensional space ($\rho$) to a scalar value ($\macro{\mu}$), we use reverse-mode AD to compute the sensitivities. For each update of the density field, we use the AD-enhanced FFT framework to calculate the homogenized stiffness tensor $\macro{\mathbb{C}}$ of the microstructure for a macroscopic strain of $\macro{\smallstrain} = [0, 0, 1.]$ under plane-stress conditions, as discussed in \Cref{sec:homogenization}. 

The optimized microstructure (see \Cref{fig:sensitivity-analysis}b)  and the converged shear modulus $\macro{\mu} \approx 0.075$ match well the results from the literature~\cite{amstutz_topological_2010, xia_design_2015, chandrasekhar_auto_2021}. Thus, integrating AD-enhanced computation of homogenized properties with AD-derived sensitivities streamlines the implementation of optimization problems. This approach allows for the simultaneous calculation of all homogenized components, differing from previous studies that required solving the static equilibrium for three separate test strains~\cite{chen_fft-based_2022}. Furthermore, integrating simultaneous homogenization of the microstructure and computation of sensitivities within an FFT-based framework facilitates the optimization of computationally intensive cases.

\section{Discussion \& Outlook}
\label{sec:discussion}
The integration of automatic differentiation (AD) with Fast-Fourier Transform (FFT) techniques can considerably simplify the numerical implementation of complex solid mechanics problems. A key advantage is the ease of deriving stress and tangent operators, thereby minimizing human error and enhancing precision. This work combines automatic differentiation with the Fourier-Galerkin method, whose variational structure facilitates straightforward integration of automatic differentiation. Here, we investigated the potential of automatic differentiation in FFT-based methods for solid mechanics, demonstrating how the automatic derivation of stress and tangent operators enables modular FFT approaches. This strengthens the use of FFT-based methods in more advanced models, particularly in multiscale scenarios where computing homogenized tangent stiffness has traditionally been a significant challenge. While we have emphasized potential applications in \Cref{sec:applicatons}, several other research areas could also greatly benefit from combining automatic differentiation with FFT-based methods.

A notable application in which we expect AD-based FFT-based methods to be beneficial is computational inelasticity. As shown in \Cref{sec:energy-derivation}, automatic differentiation greatly simplifies the extraction of the local tangent stiffness for inelastic models and is computationally equivalent to analytical expressions. Therefore, one can leverage automatic differentiation for complex plasticity models such as multi-yield surface $J_2$-plasticity~\cite{gu_consistent_2011}, Simo-plasticity~\cite{simo_computational_1998} or crystal plasticity~\cite{lucarini_fft-based_2023, takaki_phase-field_2007} for which extracting the tangent stiffness is intricate and often relies on numerical differentiation~\cite{meier_determination_2016}. Automatic differentiation can also streamline the return-mapping algorithm by boosting the robustness of trial-state projections to the yield surface~\cite{belytschko_nonlinear_2014, simo_computational_1998}. Furthermore, multiphysical problems (or coupled problems) offer another promising avenue for the application of AD-enhanced FFT techniques~\cite{pundir_fft-based_2023, sharma_fft-based_2020, chen_fft_2019}. For example, automatic differentiation can support the creation of coupled or staggered operators, leading to more precise and efficient solutions for such problems. As discussed in \cite{rothe_automatic_2015}, linearizing a thermo-mechanical problem requires four-tangent stiffness matrices; using automatic differentiation, one can greatly simplify this number to 1 or 2. Additionally, expressing them as operators can significantly reduce the memory footprint. We demonstrated that employing AD-derived operators for the St.~Venant-Kirchhoff material model, which has a straightforward analytical form (with only a few mathematical operations) for tangent stiffness, can halve memory usage.

We showed that automatic differentiation capabilities go beyond just computing stress and tangent values in an FFT-based framework. In \Cref{sec:sensitivity-analysis}, we utilized an AD-enhanced FFT-based method for topology optimization, where an accurate and efficient computation of sensitivity is necessary for a correct and fast optimization. Automatic differentiation simplifies the calculation of the sensitivity by directly differentiating the objective function with respect to the input parameters, thus eliminating the need for adjoint calculations~\cite{jodicke_efficient_2022}. Likewise, the high computational cost associated with uncertainty quantification, often due to the employed Monte Carlo approach, can be significantly reduced by using higher-order approximations to calculate expected values and variances of an output parameter. Automatic differentiation enables computing higher-order derivatives in a single simulation, thereby significantly reducing the computational cost. Furthermore, by broadening AD-FFT techniques to encompass second-order homogenization and strain gradient models, one can boost the precision and predictive power of multiscale simulations, thereby expanding the limits of computational mechanics and materials science. This adaptability highlights the potential of automatic differentiation in enhancing FFT-based methods for numerous mechanical applications.

The biggest challenge in adding AD techniques to existing codes/libraries is selecting the appropriate AD framework. Many AD frameworks are language-specific, so it is often easier to do so if the code is already in that language. For example, transforming existing Python-based code to work with differentiable frameworks such as JAX is generally straightforward (as is evident in the code provided at \cite{pundir_supplementary_2024}). JAX's functional programming makes it easy to fit into scientific workflows with minimal changes. In contrast, modifying C$++$ code to function with AD frameworks such as ADOL-C~\cite{griewank_algorithm_1996}, autodiff~\cite{noauthor_autodiffautodiff_2024} or Sacado~\cite{phipps_automatic_2022} typically requires more changes to handle automatic differentiation structures. However, these C$++$ libraries, built for performance, excel in tasks with high performance requirements, unlike Python-based framework such as JAX or PyTorch, which incur a slight computational overhead. Although frameworks such as ADOL-C and Sacado enable efficient automatic differentiation for basic arithmetic and algebra operations, allowing the derivation of stresses or tangent moduli from energy density functions, they lack, to the best of authors' knowledge, native support for FFT differentiation. Therefore, enabling automatic differentiation for FFT functions in these frameworks requires either custom adjoint routines or defining the chain rule for FFT operations, both of which are often complex and time-consuming. In contrast, JAX and PyTorch natively support automatic differentiation through FFTs, as shown in this paper, making them suitable for FFT-based methods. Furthermore, JAX compensates for AD overhead with its just-in-time compilation and advanced parallelization capabilities (CPUs and GPUs). Recently, frameworks such as Enzyme-AD~\cite{moses_instead_2020, moses_scalable_2022}, an experimental differentiable library that works across languages, enable automatic differentiation without altering the original code or libraries. This offers both flexibility, differentiation of FFT functions and high performance and, therefore, could be a great balance between ease of integration and computational efficiency. 

Another important factor in selecting an AD framework is the range of differentiation techniques it offers, such as forward-mode and reverse-mode. As discussed in \Cref{sec:automatic-differentiation}, both methods are crucial and can provide considerable efficiency depending on the function being differentiated. Although PyTorch provides AD support, mainly in reverse-mode, it does not offer as extensive support for both modes as JAX does (forward-mode, jvp function and linearize function). Therefore, the choice of a suitable framework for integrating automatic differentiation into existing code largely hinges on the user's specific requirements and preferences such as integration ease, flexibility, computing efficiency and support for parallelization. In this research, we opted for JAX to augment FFT-based methods with automatic differentiation. Nonetheless, it is feasible to employ any available AD framework, as the principles addressed in this paper remain applicable.

Finally, in this paper, we employed the Fourier-Galerkin method for the FFT approach; however, the AD technique, as demonstrated here, is applicable to other FFT-based methods~\cite{moulinec_numerical_1998, keshav_fft-based_2022, lucarini_dbfft_2019} as well.

\section{Conclusion}
\label{sec:conclusion}
In this paper, we investigated the use of automatic differentiation (AD) within the FFT-based framework for solid mechanics. We have shown that automatic differentiation allows the computation of stresses and tangent operators directly from the energy density functional, which makes FFT-based methods more accessible for complex hyperelastic materials. Furthermore, we have demonstrated that automatic differentiation allows differentiation of intricate tensor operations, FFT transforms, and Newton-Krylov solvers, which enable the computation of tangent operators for materials where local stiffness is not derivable from strain density, \textit{e.g.}, elastoplastic materials. The incorporation of automatic differentiation into the FFT-based framework decreases the reliance on explicit expressions of stresses and tangent operators while preserving accuracy and computational efficiency. We further employed the AD-enhanced FFT-based method to simplify the computation of homogenized stiffness tensors, a notable computational bottleneck in multiscale problems. We show that the AD-enhanced FFT-based approach is applicable beyond stress and tangent calculations, extending to problems like uncertainty quantification and topology optimization by simplifying sensitivity computations.  Our findings indicate that applying automatic differentiation within an FFT-based method is highly beneficial and greatly simplifies the implementation of complex solid mechanics problems, as is also evident by the simple code that we provide at~\cite{pundir_supplementary_2024}. We believe this research will encourage further studies using automatic differentiation within the FFT-based frameworks and will help in advancing the field of computational solid mechanics.

\section{CRediT authorship contribution statement}
\textbf{Mohit Pundir:} Conceptualization, Methodology, Formal analysis, Investigation, Software, Writing - Original Draft, Visualization;
\textbf{David S.\ Kammer:} Writing - Review \& Editing, Resources, Supervision, Funding acquisition

\section{Acknowledgements}
DSK and MP acknowledge support from the Swiss National Science Foundation under the SNSF starting grant (TMSGI2\_211655).
We thank Jan Zeman and Antoine Sanner for fruitful discussions and Alessandra Lingua for providing the scans of the lattice structure. We also thank Luca Michel for his feedback on the manuscript.

\section{Code Availability}
The code~\cite{pundir_supplementary_2024} for the simulation is written in Python and is an extension of the code provided in \cite{pundir_fft-based_2023}.

\clearpage

\newpage
\appendix

\section{Netwon-Krylov solver}
\label{app:newton-krylov}

\begin{algorithm}[H]
  \caption{Netwon-Krylov Solver}
  \begin{algorithmic}[1]  
    \State For a given macroscopic strain  $\macro{\strain}$ 
    \newline
        \State $\tensor{r}(\vector{x}) = \stress(\strain(\vector{x})^i),~ \strain(\vector{x})^i = \macro{\strain}$ \Comment{$\tensor{\sigma}(\tensor{\varepsilon}) = \mathbb{C}:\tensor{\varepsilon}$}
    \While {true}
    \State  solve : $\ifft{ \fft{\mathbb{G}} : \fft{\stress(\delta \strain)}  } = -\ifft{\fft{\mathbb{G}} : \fft{\tensor{\sigma}(\tensor{r})} }$ \Comment{Linear iterative solver}
    \State update : $\strain(\vector{x})^{i+1} = \strain(\vector{x})^{i} + \ifft{\fft{\delta \strain}}$  
    \If {$ \delta \strain(\vector{x})/\|\macro{\strain} \| <$  tol} {break}
    \Else
    \State $\tensor{r}(\vector{x}) =  \stress( \strain(\vector{x})^{i+1})$
    \EndIf
    \EndWhile
\end{algorithmic}
\label{alg:newton-krylov-solver}
\end{algorithm}

\section{Computation of incremental stresses for an elastoplastic material }
\label{app:elasto-plastic-material}
For a given total strain $\smallstrain^{t+\Delta t}$, we aim to find the stress state ($\stress$)and the cumulative plastic strain ($\epsilon_p$). The calculation of the stress state and the cumulative plastic strain for J2 plasticity using predictor-corrector approach is given as,  
\begin{align*}
\Delta \smallstrain &=  \smallstrain^{t+\Delta t} - \smallstrain^{t} \\
    \stress_\text{elas}^\text{tr} &= \mathbb{C}_\text{elas} : ( \smallstrain_ \text{elas}^{t} + \Delta \smallstrain) \\
    \sigma_\text{eq}^\text{tr} &= \sqrt{\dfrac{3}{2}\stress_\text{dev}^\text{tr} : \stress_\text{dev}^\text{tr}} \\
    \Phi &= \sigma_\text{eq}^\text{tr} - \underbrace{(\sigma_{y} + H p^{t})}_{\text{yield function}} \\
    \Delta \epsilon_\text{p} &= \dfrac{\langle \Phi \rangle_{+}}{3\mu + H }, \quad  \text{where}~\langle \star \rangle_{+} = \dfrac{1}{2}(\Phi + |\Phi|) \\
    \stress^{t+\Delta t} &= \stress_\text{elas}^\text{tr} - 2\mu\Delta \epsilon_\text{p} \dfrac{\stress_\text{dev}^{tr}}{\sigma_\text{eq}^\text{tr}} \\
    \epsilon_\text{p}^{t+\Delta t} &= \epsilon_\text{p}^{t} + \Delta \epsilon_\text{p} \\
    \smallstrain_\text{elas}^{t+\Delta t} &= \smallstrain_ \text{elas}^{t} + \Delta \smallstrain - \Delta \epsilon_\text{p} \dfrac{\stress_\text{dev}^{tr}}{\sigma_\text{eq}^\text{tr}}
\end{align*}
The above calculation is implemented as Python function that takes the total strain at time $t+ \Delta t$ and computes the total stress at time $t + \Delta t$. We then use the Jacobian-Vector product (using the jvp function of JAX) to compute the incremental stresses. The implementation for which is given as

\begin{minted}[frame=lines,
framesep=1mm,
baselinestretch=0.7,
fontsize=\footnotesize,
]{python}
# explicit implementation of a Python function to compute stresses
def compute_stresses(strain):

    # Compute the trial state stresses
    
    # evaluate yield surface, set to zero if elastic (or stress-free)
    
    # plastic multiplier, based on hardening law

    # apply return-map algorithm to project stresses to a feasible state
    
    return sigma

# computing the incremental stresses for an elasto-plastic material
dsigma = jax.jvp(compute_stresses, (prev_strain,), (dstrain,))[1]

\end{minted}

\section{Computation of homogenized properties for architected material using FE }
\label{app:lattice-material}

We perform a full-field computational homogenization using Finite Element method. Unlike, FFT method where we have periodic boundary condition, here we apply Kinematic Uniform Boundary Conditions (KUBC) for computational homogenization~\cite{miehe_computational_2002, kuts_computational_2024}. It assumes that the strain field is uniform on the boundary of the RVE. For a homogenized material the strain, strain relation in Voigt notation is given as $
\macro{\stress}  = \macro{\mathbb{C}} : \macro{\smallstrain} $. We express the macroscopic strain as a Identity tensor such that:
\begin{align}
 \overline{\boldsymbol{\varepsilon}} =  \begin{bmatrix} \varepsilon_{xx} & 0 & 0 \\  0 & \varepsilon_{yy} & 0   \\  0 & 0 & 2 \varepsilon_{xy} \end{bmatrix} =  \boldsymbol{I}
 \end{align}
This means that under such condition the macroscopic compliance matrix will be equal to average stress. Therefore, 
\begin{align}
 \macro{\mathbb{C}} = \begin{bmatrix} C_{xxxx} &  & \\ & C_{yyyy} & \\ & & C_{xyxy} \end{bmatrix} = \begin{bmatrix} \overline{\sigma}_{xx} &  & \\ &  \overline{\sigma}_{yy} &  \\ & &  \overline{\sigma}_{xy} \end{bmatrix}
 \end{align}
To compute the values of $\overline{\sigma}_{xx}, \overline{\sigma}_{yy}$ and $ \overline{\sigma}_{xy}$, we perform 3 FE simulations with 3 different boundary conditions. The microscale stresses for each boundary conditions are then average over the volume of the RVE as $\overline{\sigma}_{ij} = \int  \sigma_{ij}(x) dV /\Omega_\text{RVE} $. For the 2D case as considered in \Cref{sec:homogenization}, we only need 3 boundary conditions: uniaxial strain along $y$-axis, uniaxial strain along $x$-axis andd simple shear strain. We used FEniCSx for the Finite Element solution. 

\clearpage

\newpage


\begin{thebibliography}{64}
\providecommand{\natexlab}[1]{#1}
\providecommand{\url}[1]{\texttt{#1}}
\expandafter\ifx\csname urlstyle\endcsname\relax
  \providecommand{\doi}[1]{doi: #1}\else
  \providecommand{\doi}{doi: \begingroup \urlstyle{rm}\Url}\fi

\bibitem[Moulinec and Suquet(1998)]{moulinec_numerical_1998}
H.~Moulinec and P.~Suquet.
\newblock A numerical method for computing the overall response of nonlinear
  composites with complex microstructure.
\newblock \emph{Computer Methods in Applied Mechanics and Engineering},
  157\penalty0 (1):\penalty0 69--94, April 1998.
\newblock ISSN 0045-7825.
\newblock \doi{10.1016/S0045-7825(97)00218-1}.
\newblock URL
  \url{https://www.sciencedirect.com/science/article/pii/S0045782597002181}.

\bibitem[Michel et~al.(2001)Michel, Moulinec, and
  Suquet]{michel_computational_2001}
J.~C. Michel, H.~Moulinec, and P.~Suquet.
\newblock A computational scheme for linear and non-linear composites with
  arbitrary phase contrast.
\newblock \emph{International Journal for Numerical Methods in Engineering},
  52\penalty0 (1-2):\penalty0 139--160, 2001.
\newblock ISSN 1097-0207.
\newblock \doi{10.1002/nme.275}.
\newblock URL \url{https://onlinelibrary.wiley.com/doi/abs/10.1002/nme.275}.

\bibitem[Vondřejc et~al.(2014)Vondřejc, Zeman, and
  Marek]{vondrejc_fft-based_2014}
Jaroslav Vondřejc, Jan Zeman, and Ivo Marek.
\newblock An {FFT}-based {Galerkin} method for homogenization of periodic
  media.
\newblock \emph{Computers \& Mathematics with Applications}, 68\penalty0
  (3):\penalty0 156--173, August 2014.
\newblock ISSN 0898-1221.
\newblock \doi{10.1016/j.camwa.2014.05.014}.
\newblock URL
  \url{https://www.sciencedirect.com/science/article/pii/S0898122114002077}.

\bibitem[Zeman et~al.(2017)Zeman, de~Geus, Vondřejc, Peerlings, and
  Geers]{zeman_finite_2017}
J.~Zeman, T.~W.~J. de~Geus, J.~Vondřejc, R.~H.~J. Peerlings, and M.~G.~D.
  Geers.
\newblock A finite element perspective on nonlinear {FFT}-based micromechanical
  simulations.
\newblock \emph{International Journal for Numerical Methods in Engineering},
  111\penalty0 (10):\penalty0 903--926, 2017.
\newblock ISSN 1097-0207.
\newblock \doi{10.1002/nme.5481}.
\newblock URL \url{https://onlinelibrary.wiley.com/doi/abs/10.1002/nme.5481}.

\bibitem[Schneider(2021)]{schneider_review_2021}
Matti Schneider.
\newblock A review of nonlinear {FFT}-based computational homogenization
  methods.
\newblock \emph{Acta Mechanica}, 232\penalty0 (6):\penalty0 2051--2100, June
  2021.
\newblock ISSN 0001-5970, 1619-6937.
\newblock \doi{10.1007/s00707-021-02962-1}.
\newblock URL \url{https://link.springer.com/10.1007/s00707-021-02962-1}.

\bibitem[Ladecký et~al.(2023)Ladecký, Leute, Falsafi, Pultarová, Pastewka,
  Junge, and Zeman]{ladecky_optimal_2023}
Martin Ladecký, Richard~J. Leute, Ali Falsafi, Ivana Pultarová, Lars
  Pastewka, Till Junge, and Jan Zeman.
\newblock An optimal preconditioned {FFT}-accelerated finite element solver for
  homogenization.
\newblock \emph{Applied Mathematics and Computation}, 446:\penalty0 127835,
  June 2023.
\newblock ISSN 0096-3003.
\newblock \doi{10.1016/j.amc.2023.127835}.
\newblock URL
  \url{https://www.sciencedirect.com/science/article/pii/S0096300323000048}.

\bibitem[de~Geus et~al.(2017)de~Geus, Vondřejc, Zeman, Peerlings, and
  Geers]{de_geus_finite_2017}
T.~W.~J. de~Geus, J.~Vondřejc, J.~Zeman, R.~H.~J. Peerlings, and M.~G.~D.
  Geers.
\newblock Finite strain {FFT}-based non-linear solvers made simple.
\newblock \emph{Computer Methods in Applied Mechanics and Engineering},
  318:\penalty0 412--430, May 2017.
\newblock ISSN 0045-7825.
\newblock \doi{10.1016/j.cma.2016.12.032}.
\newblock URL
  \url{https://www.sciencedirect.com/science/article/pii/S0045782516318709}.

\bibitem[Rothe and Hartmann(2015)]{rothe_automatic_2015}
Steffen Rothe and Stefan Hartmann.
\newblock Automatic differentiation for stress and consistent tangent
  computation.
\newblock \emph{Archive of Applied Mechanics}, 85\penalty0 (8):\penalty0
  1103--1125, August 2015.
\newblock ISSN 1432-0681.
\newblock \doi{10.1007/s00419-014-0939-6}.
\newblock URL \url{https://doi.org/10.1007/s00419-014-0939-6}.

\bibitem[Rambausek et~al.(2019)Rambausek, Göküzüm, Nguyen, and
  Keip]{rambausek_two-scale_2019}
Matthias Rambausek, Felix~Selim Göküzüm, Lu~Trong~Khiem Nguyen, and
  Marc-André Keip.
\newblock A two-scale {FE}-{FFT} approach to nonlinear magneto-elasticity.
\newblock \emph{International Journal for Numerical Methods in Engineering},
  117\penalty0 (11):\penalty0 1117--1142, 2019.
\newblock ISSN 1097-0207.
\newblock \doi{10.1002/nme.5993}.
\newblock URL \url{https://onlinelibrary.wiley.com/doi/abs/10.1002/nme.5993}.

\bibitem[Gierden et~al.(2021)Gierden, Kochmann, Waimann, Kinner-Becker,
  Sölter, Svendsen, and Reese]{gierden_efficient_2021}
Christian Gierden, Julian Kochmann, Johanna Waimann, Tobias Kinner-Becker, Jens
  Sölter, Bob Svendsen, and Stefanie Reese.
\newblock Efficient two-scale {FE}-{FFT}-based mechanical process simulation of
  elasto-viscoplastic polycrystals at finite strains.
\newblock \emph{Computer Methods in Applied Mechanics and Engineering},
  374:\penalty0 113566, February 2021.
\newblock ISSN 0045-7825.
\newblock \doi{10.1016/j.cma.2020.113566}.
\newblock URL
  \url{https://www.sciencedirect.com/science/article/pii/S0045782520307519}.

\bibitem[Kochmann et~al.(2018)Kochmann, Ehle, Wulfinghoff, Mayer, Svendsen, and
  Reese]{kochmann_efficient_2018}
Julian Kochmann, Lisa Ehle, Stephan Wulfinghoff, Joachim Mayer, Bob Svendsen,
  and Stefanie Reese.
\newblock Efficient {Multiscale} {FE}-{FFT}-{Based} {Modeling} and {Simulation}
  of {Macroscopic} {Deformation} {Processes} with {Non}-linear {Heterogeneous}
  {Microstructures}.
\newblock In Jurica Sorić, Peter Wriggers, and Olivier Allix, editors,
  \emph{Multiscale {Modeling} of {Heterogeneous} {Structures}}, pages 129--146.
  Springer International Publishing, Cham, 2018.
\newblock ISBN 978-3-319-65463-8.
\newblock \doi{10.1007/978-3-319-65463-8_7}.
\newblock URL \url{https://doi.org/10.1007/978-3-319-65463-8_7}.

\bibitem[Felder et~al.()Felder, Kochmann, Wulﬁnghoﬀ, and
  Reese]{felder_multiscale_nodate}
Sebastian Felder, Julian Kochmann, Stephan Wulﬁnghoﬀ, and Stefanie Reese.
\newblock Multiscale {FE}-{FFT}-based thermo-mechanically coupled modeling of
  viscoplastic polycrystalline materials.

\bibitem[Tran-Duc et~al.(2024)Tran-Duc, Bunder, and
  Roberts]{tran-duc_efficient_2024}
Thien Tran-Duc, J.~E. Bunder, and A.~J. Roberts.
\newblock Efficient computational homogenisation of {2D} beams of heterogeneous
  elasticity using the patch scheme.
\newblock \emph{International Journal of Solids and Structures}, 292:\penalty0
  112719, April 2024.
\newblock ISSN 0020-7683.
\newblock \doi{10.1016/j.ijsolstr.2024.112719}.
\newblock URL
  \url{https://www.sciencedirect.com/science/article/pii/S0020768324000763}.

\bibitem[Gierden et~al.(2022)Gierden, Kochmann, Waimann, Svendsen, and
  Reese]{gierden_review_2022}
Christian Gierden, Julian Kochmann, Johanna Waimann, Bob Svendsen, and Stefanie
  Reese.
\newblock A {Review} of {FE}-{FFT}-{Based} {Two}-{Scale} {Methods} for
  {Computational} {Modeling} of {Microstructure} {Evolution} and {Macroscopic}
  {Material} {Behavior}.
\newblock \emph{Archives of Computational Methods in Engineering}, 29\penalty0
  (6):\penalty0 4115--4135, October 2022.
\newblock ISSN 1134-3060, 1886-1784.
\newblock \doi{10.1007/s11831-022-09735-6}.
\newblock URL \url{https://link.springer.com/10.1007/s11831-022-09735-6}.

\bibitem[Margossian(2019)]{margossian_review_2019}
Charles~C. Margossian.
\newblock A review of automatic differentiation and its efficient
  implementation.
\newblock \emph{WIREs Data Mining and Knowledge Discovery}, 9\penalty0
  (4):\penalty0 e1305, 2019.
\newblock ISSN 1942-4795.
\newblock \doi{10.1002/widm.1305}.
\newblock URL \url{https://onlinelibrary.wiley.com/doi/abs/10.1002/widm.1305}.

\bibitem[Vigliotti and Auricchio(2021)]{vigliotti_automatic_2021}
Andrea Vigliotti and Ferdinando Auricchio.
\newblock Automatic {Differentiation} for {Solid} {Mechanics}.
\newblock \emph{Archives of Computational Methods in Engineering}, 28\penalty0
  (3):\penalty0 875--895, May 2021.
\newblock ISSN 1886-1784.
\newblock \doi{10.1007/s11831-019-09396-y}.
\newblock URL \url{https://doi.org/10.1007/s11831-019-09396-y}.

\bibitem[Baratta et~al.(2023)Baratta, Dean, Dokken, Habera, Hale, Richardson,
  Rognes, Scroggs, Sime, and Wells]{baratta_dolfinx_2023}
Igor~A. Baratta, Joseph~P. Dean, Jørgen~S. Dokken, Michal Habera, Jack~S.
  Hale, Chris~N. Richardson, Marie~E. Rognes, Matthew~W. Scroggs, Nathan Sime,
  and Garth~N. Wells.
\newblock {DOLFINx}: the next generation {FEniCS} problem solving environment,
  2023.
\newblock Published: preprint.

\bibitem[Ansel et~al.(2024)Ansel, Yang, He, Gimelshein, Jain, Voznesensky, Bao,
  Bell, Berard, Burovski, Chauhan, Chourdia, Constable, Desmaison, DeVito,
  Ellison, Feng, Gong, Gschwind, Hirsh, Huang, Kalambarkar, Kirsch, Lazos,
  Lezcano, Liang, Liang, Lu, Luk, Maher, Pan, Puhrsch, Reso, Saroufim,
  Siraichi, Suk, Suo, Tillet, Wang, Wang, Wen, Zhang, Zhao, Zhou, Zou, Mathews,
  Chanan, Wu, and Chintala]{ansel_pytorch_2024}
Jason Ansel, Edward Yang, Horace He, Natalia Gimelshein, Animesh Jain, Michael
  Voznesensky, Bin Bao, Peter Bell, David Berard, Evgeni Burovski, Geeta
  Chauhan, Anjali Chourdia, Will Constable, Alban Desmaison, Zachary DeVito,
  Elias Ellison, Will Feng, Jiong Gong, Michael Gschwind, Brian Hirsh, Sherlock
  Huang, Kshiteej Kalambarkar, Laurent Kirsch, Michael Lazos, Mario Lezcano,
  Yanbo Liang, Jason Liang, Yinghai Lu, CK~Luk, Bert Maher, Yunjie Pan,
  Christian Puhrsch, Matthias Reso, Mark Saroufim, Marcos~Yukio Siraichi, Helen
  Suk, Michael Suo, Phil Tillet, Eikan Wang, Xiaodong Wang, William Wen,
  Shunting Zhang, Xu~Zhao, Keren Zhou, Richard Zou, Ajit Mathews, Gregory
  Chanan, Peng Wu, and Soumith Chintala.
\newblock {PyTorch} 2: {Faster} {Machine} {Learning} {Through} {Dynamic}
  {Python} {Bytecode} {Transformation} and {Graph} {Compilation}, April 2024.
\newblock URL \url{https://pytorch.org/assets/pytorch2-2.pdf}.
\newblock Publication Title: 29th ACM International Conference on Architectural
  Support for Programming Languages and Operating Systems, Volume 2 (ASPLOS
  '24) original-date: 2016-08-13T05:26:41Z.

\bibitem[Bradbury et~al.(2018)Bradbury, Frostig, Hawkins, Johnson, Leary,
  Maclaurin, Necula, Paszke, VanderPlas, Wanderman-Milne, and
  Zhang]{bradbury_jax_2018}
James Bradbury, Roy Frostig, Peter Hawkins, Matthew~James Johnson, Chris Leary,
  Dougal Maclaurin, George Necula, Adam Paszke, Jake VanderPlas, Skye
  Wanderman-Milne, and Qiao Zhang.
\newblock {JAX}: composable transformations of {Python}+{NumPy} programs, 2018.
\newblock URL \url{http://github.com/google/jax}.

\bibitem[Schoenholz and Cubuk(2020)]{schoenholz_jax_2020}
Samuel~S. Schoenholz and Ekin~D. Cubuk.
\newblock {JAX}, {M}.{D}.: {A} {Framework} for {Differentiable} {Physics},
  December 2020.
\newblock URL \url{http://arxiv.org/abs/1912.04232}.
\newblock arXiv:1912.04232 [cond-mat, physics:physics, stat].

\bibitem[Carrer et~al.(2024)Carrer, Cezar, Bore, Ledum, and
  Cascella]{carrer_learning_2024}
Manuel Carrer, Henrique~Musseli Cezar, Sigbjørn~Løland Bore, Morten Ledum,
  and Michele Cascella.
\newblock Learning {Force} {Field} {Parameters} from {Differentiable}
  {Particle}-{Field} {Molecular} {Dynamics}.
\newblock \emph{Journal of Chemical Information and Modeling}, July 2024.
\newblock ISSN 1549-9596.
\newblock \doi{10.1021/acs.jcim.4c00564}.
\newblock URL \url{https://doi.org/10.1021/acs.jcim.4c00564}.
\newblock Publisher: American Chemical Society.

\bibitem[Toshev et~al.(2024)Toshev, Ramachandran, Erbesdobler, Galletti,
  Brandstetter, and Adams]{toshev_jax-sph_2024}
Artur~P. Toshev, Harish Ramachandran, Jonas~A. Erbesdobler, Gianluca Galletti,
  Johannes Brandstetter, and Nikolaus~A. Adams.
\newblock {JAX}-{SPH}: {A} {Differentiable} {Smoothed} {Particle}
  {Hydrodynamics} {Framework}, March 2024.
\newblock URL \url{http://arxiv.org/abs/2403.04750}.
\newblock arXiv:2403.04750 [physics].

\bibitem[Xue et~al.(2023)Xue, Liao, Gan, Park, Xie, Liu, and
  Cao]{xue_jax-fem_2023}
Tianju Xue, Shuheng Liao, Zhengtao Gan, Chanwook Park, Xiaoyu Xie, Wing~Kam
  Liu, and Jian Cao.
\newblock {JAX}-{FEM}: {A} differentiable {GPU}-accelerated {3D} finite element
  solver for automatic inverse design and mechanistic data science.
\newblock \emph{Computer Physics Communications}, 291:\penalty0 108802, October
  2023.
\newblock ISSN 0010-4655.
\newblock \doi{10.1016/j.cpc.2023.108802}.
\newblock URL
  \url{https://www.sciencedirect.com/science/article/pii/S0010465523001479}.

\bibitem[Bleyer(2024)]{bleyer_numerical_2024}
Jeremy Bleyer.
\newblock Numerical tours of {Computational} {Mechanics} with {FEniCSx},
  January 2024.
\newblock URL \url{https://doi.org/10.5281/zenodo.10470942}.

\bibitem[Blühdorn et~al.(2022)Blühdorn, Gauger, and
  Kabel]{bluhdorn_automat_2022}
Johannes Blühdorn, Nicolas~R. Gauger, and Matthias Kabel.
\newblock {AutoMat}: automatic differentiation for generalized standard
  materials on {GPUs}.
\newblock \emph{Computational Mechanics}, 69\penalty0 (2):\penalty0 589--613,
  February 2022.
\newblock ISSN 1432-0924.
\newblock \doi{10.1007/s00466-021-02105-2}.
\newblock URL \url{https://doi.org/10.1007/s00466-021-02105-2}.

\bibitem[Inc()]{inc_mathematica_nodate}
Wolfram~Research Inc.
\newblock Mathematica, {Version} 14.1.
\newblock URL \url{https://www.wolfram.com/mathematica}.

\bibitem[Pundir et~al.(2023)Pundir, Kammer, and Angst]{pundir_fft-based_2023}
Mohit Pundir, David~S. Kammer, and Ueli Angst.
\newblock An {FFT}-based framework for predicting corrosion-driven damage in
  fractal porous media.
\newblock \emph{Journal of the Mechanics and Physics of Solids}, 179:\penalty0
  105388, October 2023.
\newblock ISSN 00225096.
\newblock \doi{10.1016/j.jmps.2023.105388}.
\newblock URL
  \url{https://linkinghub.elsevier.com/retrieve/pii/S0022509623001928}.

\bibitem[Hirsch et~al.(2013)Hirsch, Smale, and
  Devaney]{hirsch_differential_2013}
M.~W. Hirsch, S.~Smale, and R.~L. Devaney.
\newblock \emph{Differential {Equations}, {Dynamical} {Systems}, and an
  {Introduction} to {Chaos}}.
\newblock Elsevier, 3 edition, 2013.
\newblock ISBN 978-0-12-382010-5.
\newblock URL \url{https://linkinghub.elsevier.com/retrieve/pii/C20090611600}.

\bibitem[Balestriero and Baraniuk(2021)]{balestriero_fast_2021}
Randall Balestriero and Richard Baraniuk.
\newblock Fast {Jacobian}-{Vector} {Product} for {Deep} {Networks}, March 2021.
\newblock URL \url{http://arxiv.org/abs/2104.00219}.
\newblock arXiv:2104.00219 [cs].

\bibitem[Eshelby and Peierls(1997)]{eshelby_determination_1997}
John~Douglas Eshelby and Rudolf~Ernst Peierls.
\newblock The determination of the elastic field of an ellipsoidal inclusion,
  and related problems.
\newblock \emph{Proceedings of the Royal Society of London. Series A.
  Mathematical and Physical Sciences}, 241\penalty0 (1226):\penalty0 376--396,
  January 1997.
\newblock \doi{10.1098/rspa.1957.0133}.
\newblock URL
  \url{https://royalsocietypublishing.org/doi/10.1098/rspa.1957.0133}.
\newblock Publisher: Royal Society.

\bibitem[Blanc and Le~Bris(2023)]{blanc_homogenization_2023}
Xavier Blanc and Claude Le~Bris.
\newblock \emph{Homogenization {Theory} for {Multiscale} {Problems}: {An}
  introduction}, volume~21 of \emph{{MS}\&{A}}.
\newblock Springer Nature Switzerland, Cham, 2023.
\newblock ISBN 978-3-031-21832-3 978-3-031-21833-0.
\newblock \doi{10.1007/978-3-031-21833-0}.
\newblock URL \url{https://link.springer.com/10.1007/978-3-031-21833-0}.

\bibitem[Geers et~al.(2010)Geers, Kouznetsova, and
  Brekelmans]{geers_multi-scale_2010}
M.~G.~D. Geers, V.~G. Kouznetsova, and W.~A.~M. Brekelmans.
\newblock Multi-scale computational homogenization: {Trends} and challenges.
\newblock \emph{Journal of Computational and Applied Mathematics}, 234\penalty0
  (7):\penalty0 2175--2182, August 2010.
\newblock ISSN 0377-0427.
\newblock \doi{10.1016/j.cam.2009.08.077}.
\newblock URL
  \url{https://www.sciencedirect.com/science/article/pii/S0377042709005536}.

\bibitem[Meier et~al.(2016)Meier, Schwarz, and
  Werner]{meier_determination_2016}
Felix Meier, Cornelia Schwarz, and Ewald Werner.
\newblock Determination of the tangent stiffness tensor in materials modeling
  in case of large deformations by calculation of a directed strain
  perturbation.
\newblock \emph{Computer Methods in Applied Mechanics and Engineering},
  300:\penalty0 628--642, March 2016.
\newblock ISSN 0045-7825.
\newblock \doi{10.1016/j.cma.2015.11.034}.
\newblock URL
  \url{https://www.sciencedirect.com/science/article/pii/S0045782515003989}.

\bibitem[Göküzüm and Keip(2018)]{gokuzum_algorithmically_2018}
Felix~Selim Göküzüm and Marc-André Keip.
\newblock An algorithmically consistent macroscopic tangent operator for
  {FFT}-based computational homogenization.
\newblock \emph{International Journal for Numerical Methods in Engineering},
  113\penalty0 (4):\penalty0 581--600, 2018.
\newblock ISSN 1097-0207.
\newblock \doi{10.1002/nme.5627}.
\newblock URL \url{https://onlinelibrary.wiley.com/doi/abs/10.1002/nme.5627}.

\bibitem[Pundir and Kammer(2024)]{pundir_supplementary_2024}
Mohit Pundir and David~S. Kammer.
\newblock Supplementary {Code} - {Simplifying} {FFT}-based methods for
  mechanics with automatic differentiation, 2024.
\newblock URL
  \url{https://gitlab.ethz.ch/cmbm-public/papers-supp-info/2024/simplifying-fft-based-methods-for-mechanics-with-automatic-differentiation}.

\bibitem[Simo and Hughes(1998)]{simo_computational_1998}
J.~C. Simo and T.~J.~R. Hughes.
\newblock \emph{Computational {Inelasticity}}, volume~7.
\newblock Springer-Verlag, New York, 1998.
\newblock ISBN 978-0-387-97520-7.
\newblock \doi{10.1007/b98904}.
\newblock URL \url{http://link.springer.com/10.1007/b98904}.

\bibitem[Belytschko et~al.(2014)Belytschko, Liu, Moran, and
  Elkhodary]{belytschko_nonlinear_2014}
Ted Belytschko, W.~K. Liu, B.~Moran, and K.~I. Elkhodary.
\newblock \emph{Nonlinear {Finite} {Elements} for {Continua} and {Structures}}.
\newblock Wiley, 2 edition, 2014.

\bibitem[Balan~P et~al.(2023)Balan~P, Mertens~A, and
  Bahubalendruni]{balan_p_auxetic_2023}
Madhu Balan~P, Johnney Mertens~A, and M~V A~Raju Bahubalendruni.
\newblock Auxetic mechanical metamaterials and their futuristic developments:
  {A} state-of-art review.
\newblock \emph{Materials Today Communications}, 34:\penalty0 105285, March
  2023.
\newblock ISSN 2352-4928.
\newblock \doi{10.1016/j.mtcomm.2022.105285}.
\newblock URL
  \url{https://www.sciencedirect.com/science/article/pii/S2352492822021262}.

\bibitem[Karapiperis and Kochmann(2023)]{karapiperis_prediction_2023}
Konstantinos Karapiperis and Dennis~M. Kochmann.
\newblock Prediction and control of fracture paths in disordered architected
  materials using graph neural networks.
\newblock \emph{Communications Engineering}, 2\penalty0 (1):\penalty0 1--9,
  June 2023.
\newblock ISSN 2731-3395.
\newblock \doi{10.1038/s44172-023-00085-0}.
\newblock URL \url{https://www.nature.com/articles/s44172-023-00085-0}.
\newblock Publisher: Nature Publishing Group.

\bibitem[Magrini et~al.(2024)Magrini, Fox, Wihardja, Kolli, and
  Daraio]{magrini_control_2024}
Tommaso Magrini, Chelsea Fox, Adeline Wihardja, Athena Kolli, and Chiara
  Daraio.
\newblock Control of {Mechanical} and {Fracture} {Properties} in {Two}-{Phase}
  {Materials} {Reinforced} by {Continuous}, {Irregular} {Networks}.
\newblock \emph{Advanced Materials}, 36\penalty0 (6):\penalty0 2305198, 2024.
\newblock ISSN 1521-4095.
\newblock \doi{10.1002/adma.202305198}.
\newblock URL
  \url{https://onlinelibrary.wiley.com/doi/abs/10.1002/adma.202305198}.

\bibitem[Shaikeea et~al.(2022)Shaikeea, Cui, O’Masta, Zheng, and
  Deshpande]{shaikeea_toughness_2022}
Angkur Jyoti~Dipanka Shaikeea, Huachen Cui, Mark O’Masta, Xiaoyu~Rayne Zheng,
  and Vikram~Sudhir Deshpande.
\newblock The toughness of mechanical metamaterials.
\newblock \emph{Nature Materials}, 21\penalty0 (3):\penalty0 297--304, March
  2022.
\newblock ISSN 1476-4660.
\newblock \doi{10.1038/s41563-021-01182-1}.
\newblock URL \url{https://www.nature.com/articles/s41563-021-01182-1}.
\newblock Publisher: Nature Publishing Group.

\bibitem[Miehe and Koch(2002)]{miehe_computational_2002}
C.~Miehe and A.~Koch.
\newblock Computational micro-to-macro transitions of discretized
  microstructures undergoing small strains.
\newblock \emph{Archive of Applied Mechanics}, 72\penalty0 (4):\penalty0
  300--317, July 2002.
\newblock ISSN 1432-0681.
\newblock \doi{10.1007/s00419-002-0212-2}.
\newblock URL \url{https://doi.org/10.1007/s00419-002-0212-2}.

\bibitem[Zeman et~al.(2010)Zeman, Vondřejc, Novák, and
  Marek]{zeman_accelerating_2010}
J.~Zeman, J.~Vondřejc, J.~Novák, and I.~Marek.
\newblock Accelerating a {FFT}-based solver for numerical homogenization of
  periodic media by conjugate gradients.
\newblock \emph{Journal of Computational Physics}, 229\penalty0 (21):\penalty0
  8065--8071, October 2010.
\newblock ISSN 00219991.
\newblock \doi{10.1016/j.jcp.2010.07.010}.
\newblock URL \url{http://arxiv.org/abs/1004.1122}.
\newblock arXiv:1004.1122 [cond-mat, physics:physics].

\bibitem[Bendsøe and Sigmund(2004)]{bendsoe_topology_2004}
Martin~P. Bendsøe and Ole Sigmund.
\newblock \emph{Topology {Optimization}}.
\newblock Springer, Berlin, Heidelberg, 2004.
\newblock ISBN 978-3-642-07698-5 978-3-662-05086-6.
\newblock \doi{10.1007/978-3-662-05086-6}.
\newblock URL \url{http://link.springer.com/10.1007/978-3-662-05086-6}.

\bibitem[Svanberg(1987)]{svanberg_method_1987}
Krister Svanberg.
\newblock The method of moving asymptotes—a new method for structural
  optimization.
\newblock \emph{International Journal for Numerical Methods in Engineering},
  24\penalty0 (2):\penalty0 359--373, 1987.
\newblock ISSN 1097-0207.
\newblock \doi{10.1002/nme.1620240207}.
\newblock URL
  \url{https://onlinelibrary.wiley.com/doi/abs/10.1002/nme.1620240207}.

\bibitem[Chandrasekhar et~al.(2021)Chandrasekhar, Sridhara, and
  Suresh]{chandrasekhar_auto_2021}
Aaditya Chandrasekhar, Saketh Sridhara, and Krishnan Suresh.
\newblock {AuTO}: {A} {Framework} for {Automatic} differentiation in {Topology}
  {Optimization}.
\newblock \emph{Structural and Multidisciplinary Optimization}, 64\penalty0
  (6):\penalty0 4355--4365, December 2021.
\newblock ISSN 1615-147X, 1615-1488.
\newblock \doi{10.1007/s00158-021-03025-8}.
\newblock URL \url{http://arxiv.org/abs/2104.01965}.
\newblock arXiv:2104.01965 [cs, math].

\bibitem[Xia and Breitkopf(2015)]{xia_design_2015}
Liang Xia and Piotr Breitkopf.
\newblock Design of materials using topology optimization and energy-based
  homogenization approach in {Matlab}.
\newblock \emph{Structural and Multidisciplinary Optimization}, 52\penalty0
  (6):\penalty0 1229--1241, December 2015.
\newblock ISSN 1615-1488.
\newblock \doi{10.1007/s00158-015-1294-0}.
\newblock URL \url{https://doi.org/10.1007/s00158-015-1294-0}.

\bibitem[Andreassen et~al.(2011)Andreassen, Clausen, Schevenels, Lazarov, and
  Sigmund]{andreassen_efficient_2011}
Erik Andreassen, Anders Clausen, Mattias Schevenels, Boyan~S. Lazarov, and Ole
  Sigmund.
\newblock Efficient topology optimization in {MATLAB} using 88 lines of code.
\newblock \emph{Structural and Multidisciplinary Optimization}, 43\penalty0
  (1):\penalty0 1--16, January 2011.
\newblock ISSN 1615-1488.
\newblock \doi{10.1007/s00158-010-0594-7}.
\newblock URL \url{https://doi.org/10.1007/s00158-010-0594-7}.

\bibitem[Amstutz et~al.(2010)Amstutz, Giusti, Novotny, and
  de~Souza~Neto]{amstutz_topological_2010}
S.~Amstutz, S.~M. Giusti, A.~A. Novotny, and E.~A. de~Souza~Neto.
\newblock Topological derivative for multi-scale linear elasticity models
  applied to the synthesis of microstructures.
\newblock \emph{International Journal for Numerical Methods in Engineering},
  84\penalty0 (6):\penalty0 733--756, 2010.
\newblock ISSN 1097-0207.
\newblock \doi{10.1002/nme.2922}.
\newblock URL \url{https://onlinelibrary.wiley.com/doi/abs/10.1002/nme.2922}.

\bibitem[Chen et~al.(2022)Chen, Wu, Xie, Wu, and Zhou]{chen_fft-based_2022}
Zeyao Chen, Baisheng Wu, Yi~Min Xie, Xian Wu, and Shiwei Zhou.
\newblock {FFT}-based {Inverse} {Homogenization} for {Cellular} {Material}
  {Design}.
\newblock \emph{International Journal of Mechanical Sciences}, 231:\penalty0
  107572, October 2022.
\newblock ISSN 0020-7403.
\newblock \doi{10.1016/j.ijmecsci.2022.107572}.
\newblock URL
  \url{https://www.sciencedirect.com/science/article/pii/S0020740322004635}.

\bibitem[Gu et~al.(2011)Gu, Conte, Yang, and Elgamal]{gu_consistent_2011}
Q.~Gu, J.~P. Conte, Z.~Yang, and A.~Elgamal.
\newblock Consistent tangent moduli for multi-yield-surface {J2} plasticity
  model.
\newblock \emph{Computational Mechanics}, 48\penalty0 (1):\penalty0 97--120,
  July 2011.
\newblock ISSN 0178-7675, 1432-0924.
\newblock \doi{10.1007/s00466-011-0576-7}.
\newblock URL \url{http://link.springer.com/10.1007/s00466-011-0576-7}.

\bibitem[Lucarini et~al.(2023)Lucarini, Dunne, and
  Martínez-Pañeda]{lucarini_fft-based_2023}
S.~Lucarini, F.~P.~E. Dunne, and E.~Martínez-Pañeda.
\newblock An {FFT}-based crystal plasticity phase-field model for
  micromechanical fatigue cracking based on the stored energy density.
\newblock \emph{International Journal of Fatigue}, 172:\penalty0 107670, July
  2023.
\newblock ISSN 0142-1123.
\newblock \doi{10.1016/j.ijfatigue.2023.107670}.
\newblock URL
  \url{https://www.sciencedirect.com/science/article/pii/S0142112323001718}.

\bibitem[Takaki et~al.(2007)Takaki, Yamanaka, Higa, and
  Tomita]{takaki_phase-field_2007}
T.~Takaki, A.~Yamanaka, Y.~Higa, and Y.~Tomita.
\newblock Phase-field model during static recrystallization based on
  crystal-plasticity theory.
\newblock \emph{Journal of Computer-Aided Materials Design}, 14\penalty0
  (1):\penalty0 75--84, December 2007.
\newblock ISSN 1573-4900.
\newblock \doi{10.1007/s10820-007-9083-8}.
\newblock URL \url{https://doi.org/10.1007/s10820-007-9083-8}.

\bibitem[Sharma et~al.(2020)Sharma, Peerlings, Shanthraj, Roters, and
  Geers]{sharma_fft-based_2020}
L.~Sharma, R.~H.~J. Peerlings, P.~Shanthraj, F.~Roters, and M.~G.~D. Geers.
\newblock An {FFT}-based spectral solver for interface decohesion modelling
  using a gradient damage approach.
\newblock \emph{Computational Mechanics}, 65\penalty0 (4):\penalty0 925--939,
  April 2020.
\newblock ISSN 0178-7675, 1432-0924.
\newblock \doi{10.1007/s00466-019-01801-4}.
\newblock URL \url{http://link.springer.com/10.1007/s00466-019-01801-4}.

\bibitem[Chen et~al.(2019)Chen, Vasiukov, Gélébart, and Park]{chen_fft_2019}
Yang Chen, Dmytro Vasiukov, Lionel Gélébart, and Chung~Hae Park.
\newblock A {FFT} solver for variational phase-field modeling of brittle
  fracture.
\newblock \emph{Computer Methods in Applied Mechanics and Engineering},
  349:\penalty0 167--190, June 2019.
\newblock ISSN 0045-7825.
\newblock \doi{10.1016/j.cma.2019.02.017}.
\newblock URL
  \url{https://www.sciencedirect.com/science/article/pii/S004578251930088X}.

\bibitem[Jödicke et~al.(2022)Jödicke, Leute, Junge, and
  Pastewka]{jodicke_efficient_2022}
Indre Jödicke, Richard~J. Leute, Till Junge, and Lars Pastewka.
\newblock Efficient topology optimization using compatibility projection in
  micromechanical homogenization, June 2022.
\newblock URL \url{http://arxiv.org/abs/2107.04123}.
\newblock arXiv:2107.04123 [cs].

\bibitem[Griewank et~al.(1996)Griewank, Juedes, and
  Utke]{griewank_algorithm_1996}
Andreas Griewank, David Juedes, and Jean Utke.
\newblock Algorithm 755: {ADOL}-{C}: a package for the automatic
  differentiation of algorithms written in {C}/{C}++.
\newblock \emph{ACM Trans. Math. Softw.}, 22\penalty0 (2):\penalty0 131--167,
  June 1996.
\newblock ISSN 0098-3500.
\newblock \doi{10.1145/229473.229474}.
\newblock URL \url{https://dl.acm.org/doi/10.1145/229473.229474}.

\bibitem[noa(2024)]{noauthor_autodiffautodiff_2024}
autodiff/autodiff, October 2024.
\newblock URL \url{https://github.com/autodiff/autodiff}.
\newblock original-date: 2018-07-19T07:28:38Z.

\bibitem[Phipps et~al.(2022)Phipps, Pawlowski, and
  Trott]{phipps_automatic_2022}
Eric Phipps, Roger Pawlowski, and Christian Trott.
\newblock Automatic {Differentiation} of {C}++ {Codes} on {Emerging} {Manycore}
  {Architectures} with {Sacado}.
\newblock \emph{ACM Trans. Math. Softw.}, 48\penalty0 (4):\penalty0
  43:1--43:29, December 2022.
\newblock ISSN 0098-3500.
\newblock \doi{10.1145/3560262}.
\newblock URL \url{https://dl.acm.org/doi/10.1145/3560262}.

\bibitem[Moses and Churavy(2020)]{moses_instead_2020}
William Moses and Valentin Churavy.
\newblock Instead of {Rewriting} {Foreign} {Code} for {Machine} {Learning},
  {Automatically} {Synthesize} {Fast} {Gradients}.
\newblock In H.~Larochelle, M.~Ranzato, R.~Hadsell, M.~F. Balcan, and H.~Lin,
  editors, \emph{Advances in {Neural} {Information} {Processing} {Systems}},
  volume~33, pages 12472--12485. Curran Associates, Inc., 2020.
\newblock URL
  \url{https://proceedings.neurips.cc/paper/2020/file/9332c513ef44b682e9347822c2e457ac-Paper.pdf}.

\bibitem[Moses et~al.(2022)Moses, Narayanan, Paehler, Churavy, Schanen,
  Hückelheim, Doerfert, and Hovland]{moses_scalable_2022}
William~S. Moses, Sri Hari~Krishna Narayanan, Ludger Paehler, Valentin Churavy,
  Michel Schanen, Jan Hückelheim, Johannes Doerfert, and Paul Hovland.
\newblock Scalable {Automatic} {Differentiation} of {Multiple} {Parallel}
  {Paradigms} through {Compiler} {Augmentation}.
\newblock In \emph{Proceedings of the {International} {Conference} on {High}
  {Performance} {Computing}, {Networking}, {Storage} and {Analysis}}, {SC} '22.
  IEEE Press, 2022.
\newblock ISBN 9784665454445.
\newblock Place: Dallas, Texas.

\bibitem[Keshav et~al.(2022)Keshav, Fritzen, and Kabel]{keshav_fft-based_2022}
Sanath Keshav, Felix Fritzen, and Matthias Kabel.
\newblock \emph{{FFT}-based {Homogenization} at {Finite} {Strains} using
  {Composite} {Boxels} ({ComBo})}.
\newblock April 2022.

\bibitem[Lucarini and Segurado(2019)]{lucarini_dbfft_2019}
S.~Lucarini and J.~Segurado.
\newblock {DBFFT}: {A} displacement based {FFT} approach for non-linear
  homogenization of the mechanical behavior.
\newblock \emph{International Journal of Engineering Science}, 144:\penalty0
  103131, November 2019.
\newblock ISSN 0020-7225.
\newblock \doi{10.1016/j.ijengsci.2019.103131}.
\newblock URL
  \url{https://www.sciencedirect.com/science/article/pii/S0020722519304446}.

\bibitem[Kuts et~al.(2024)Kuts, Walker, and Newell]{kuts_computational_2024}
Mikhail Kuts, James Walker, and Pania Newell.
\newblock Computational homogenization of linear elastic properties in porous
  non-woven fibrous materials.
\newblock \emph{Mechanics of Materials}, 189:\penalty0 104868, February 2024.
\newblock ISSN 0167-6636.
\newblock \doi{10.1016/j.mechmat.2023.104868}.
\newblock URL
  \url{https://www.sciencedirect.com/science/article/pii/S0167663623003149}.

\end{thebibliography}
\end{document}